\documentclass{aa}  

\usepackage{graphicx}
\usepackage{txfonts}
\usepackage{amsmath,amstext}
\usepackage{hyperref}

\usepackage[figure,figure*]{hypcap}
\usepackage{appendix}
\usepackage{soul}
\usepackage{booktabs}
\usepackage{multirow}
\usepackage{rotating}
\usepackage{xcolor}
\usepackage{array}
\usepackage{longtable}
\usepackage{multirow}
\usepackage{supertabular,booktabs}
\usepackage{multicol}
\usepackage{url}

\usepackage{graphicx}
\usepackage{txfonts}
\begin{document}

\title{Photometric redshifts from SDSS images using a Convolutional Neural Network}

\author{Johanna Pasquet\inst{\ref{inst1}}, E. Bertin\inst{\ref{inst2}}, M. Treyer \inst{\ref{inst3}},  S. Arnouts\inst{\ref{inst3}} and D. Fouchez\inst{\ref{inst1}}}

\institute{Aix Marseille Univ, CNRS/IN2P3, CPPM, Marseille, France\label{inst1}\and
 Sorbonne Universit\'e, CNRS, UMR 7095, Institut d’Astrophysique de Paris, 98 bis bd Arago, 75014 Paris, France\label{inst2}\and
Aix Marseille Univ, CNRS, CNES, LAM, Marseille, France\label{inst3}}

   \date{}

 
  \abstract
{We developed a Deep Convolutional Neural Network (CNN), used as a classifier, to estimate photometric redshifts and associated probability distribution functions (PDF) for galaxies in the Main Galaxy Sample of the Sloan Digital Sky Survey at $z < 0.4$. Our method exploits all the information present in the images without any feature extraction. The input data consist of 64$\times$64 pixel \textit{ugriz} images centered on the spectroscopic targets, plus the galactic reddening value on the line-of-sight. For training sets of 100k objects or more ($\ge 20$\% of the database), we reach a dispersion $\sigma_{\rm MAD}<0.01$, significantly lower than the current best one obtained from another machine learning technique on the same sample. The bias is lower than $10^{-4}$, independent of photometric redshift. The PDFs are shown to have very good predictive power. We also find that the CNN redshifts are unbiased with respect to galaxy inclination, and that $\sigma_{\rm MAD}$ decreases with the signal-to-noise ratio (SNR), achieving values below $0.007$ for SNR $>100$, as in the deep stacked region of Stripe 82. We argue that for most galaxies the precision is limited by the SNR of SDSS images rather than by the method. The success of this experiment at low redshift opens promising perspectives for upcoming surveys.}

 
\titlerunning{Photometric redshifts from SDSS galaxy images}
\authorrunning{J. Pasquet et al.}

\maketitle
%
%
%
%

\section{Introduction}\label{sec:intro}

Panoramic imaging surveys for cosmology are underway or in preparation phase (HSC, LSST, Euclid, WFIRST). They will deliver multi-band photometry for billions of galaxies for which reliable redshifts are necessary to study the large scale structure of the universe and to constrain the dark energy equation-of-state using weak gravitational lensing. However spectroscopic redshifts are extremely time-intensive and it has become a necessity to use photometric redshifts. The projections for cosmic shear measurements estimate that the true mean redshift of objects in each photometric redshift bin must be known to better than $\sim$0.002(1+z) \citep[]{Knox2006} with stringent requirements on the fraction of unconstrained catastrophic outliers \citep[]{Hearin2010}. Another challenge is the derivation of robust redshift probability distribution functions \citep[PDFs,][]{Mandelbaum2008} for a complete understanding of the uncertainties attached to any measurements in cosmology (e.g. galaxy clustering, weak lensing tomography, baryon acoustic oscillations) or galaxy evolution (e.g. luminosity and stellar mass functions, galaxy density field reconstruction, cluster finders).
 
 Two main techniques are traditionally used to perform this task:  template fitting and machine learning algorithms.  The template fitting codes \citep[e.g.,][]{ Arnouts1999,Benitez2000,  Brammer2008} match the broad band photometry of a galaxy to the synthetic magnitudes of a suite of templates across a large redshift interval\footnote{Baum (\citeyear{1962IAUS...15..390B}) first developed this method by observing the spectral energy distribution of six elliptic galaxies in the Virgo cluster in nine bands from 3730$\text{\AA}$ to 9875$\text{\AA}$. }. This technique does not require a large spectroscopic sample for training: when a representative set of galaxy template has been found it can be applied to different surveys and redshift range. However, they are often computationally intensive due to the brute-force approach to explore the pre-generated grid of model photometry. Moreover poorly known parameters, such as dust attenuation, can lead to degeneracies in color $-$ redshift space.  On the other hand, the machine learning methods, such as artificial neural network \citep[]{Collister2004}, k$-$nearest neighbors \citep[KNN][]{Csabai2007} or random forest techniques \citep[]{Carliles2010} were shown to have similar or better performances when a large  spectroscopic training set is available. However, they are only reliable within the limits of the training set and the current lack of spectroscopic coverage in some color space regions and at high redshift remains a major issue for this approach. For these reasons, template-fitting methods are still considered and new approaches have emerged which combine several techniques to maximize the photometric redshift PDF estimations \citep[e.g.,][]{CarrascoKind2014, Cavuoti2017}.

One limiting factor of all the above methods is the information used as input. The accuracy of the output photometric redshifts is limited by that of the photometric measurements \citep[]{Hildebrandt2012}. Magnitudes or colors are affected by the choice of aperture size, Point Spread Function (PSF) variations, overlapping sources, and even modeled magnitudes (accounting for PSF and galaxy luminosity profiles) capture only a fraction of the information present in the images.

Over the past few years, Deep Learning techniques have revolutionized the field of image recognition. By bypassing the condensed information of manual feature extraction required by previous methods they can offer unprecedented image classification performance in a number of astronomical problems, including galaxy morphological classification \citep[e.g.,][]{Dieleman2015, HuertasCompany2015}, lensed images \citep[]{Lanusse2018}, classification of light curves \citep[]{charnock2017,Pasquet2017}. Thanks to the speed boost from Graphic Processing Units (GPU) technology and large galaxy spectroscopic sample such as the  SDSS survey, \citet[]{Hoyle2016, disanto2018} showed that Deep Convolutional neural network (CNN)  were able to provide accurate phototometric redshifts from multichannel images, instead of extracted features, taking advantage of all the information contained in the pixels, such as galaxy surface brightness and size, disk inclination, or the presence of color gradients and neighbors. To do so, \citet[]{Hoyle2016} used a Deep CNN inspired by the architecture of \citet[]{ImageNet},  on 60$\times$60 RGBA images, encoding colors ($i-z$,  $r-i$, $g-r$) in RGB layers and $r$ band magnitudes in the alpha layer. The use of only four bands ($griz$) allowed them to mimic the DES experiment. They divided the  spectroscopic redshift distribution into bins and extracted the redshift bin that the galaxy was most likely to be in. To obtain a PDF, they then randomly extracted 100 60$\times$60 stamps from an original 72 $\times$72 image stamp, and rerun the CNN algorithm. \citet[]{disanto2018} used a Deep CNN  model, based on a LeNet-5 architecture \citep[]{LeCun1998}, using 28$\times$28 pixel  images in the 5 SDSS bands ($ugriz$) as well as all the color combinations as input. They modified the fully connected part of the CNN to include a Mixture Density Network (with one hidden layer) to describe the output PDF as a Gaussian mixture model.  These two first-of-their-kind analyses based on existing architectures, achieved competitive photometric redshift accuracies compared to other machine learning techniques based on boosted decision tree or random forest.

In this paper, we present a Deep Learning model for the estimation of photometric redshifts and their associated PDF using the TensorFlow framework.  In contrast to previous studies, our input consists of $ugriz$ images only (no color images are produced), from the flux-limited  spectroscopic Main Galaxy Sample (MGS) of the SDSS. The paper is organized as follows. In Section 2, we describe the data used in this study. In Section 3, we introduce the CNN concepts and the particular CNN architecture we are proposing. In Section 4, we present our results for the estimation of photometric redshifts and associated PDFs. In Section 5, we investigate the impact of reducing the size of the training set. In Section 6, we analyze the behavior of the CNN with respect to a number of galaxy properties and observing conditions.
Our results are summarized in Section 7. 


\section{Data} \label{sec:style}
\begin{figure}[h!]
\includegraphics[width=9.3cm]{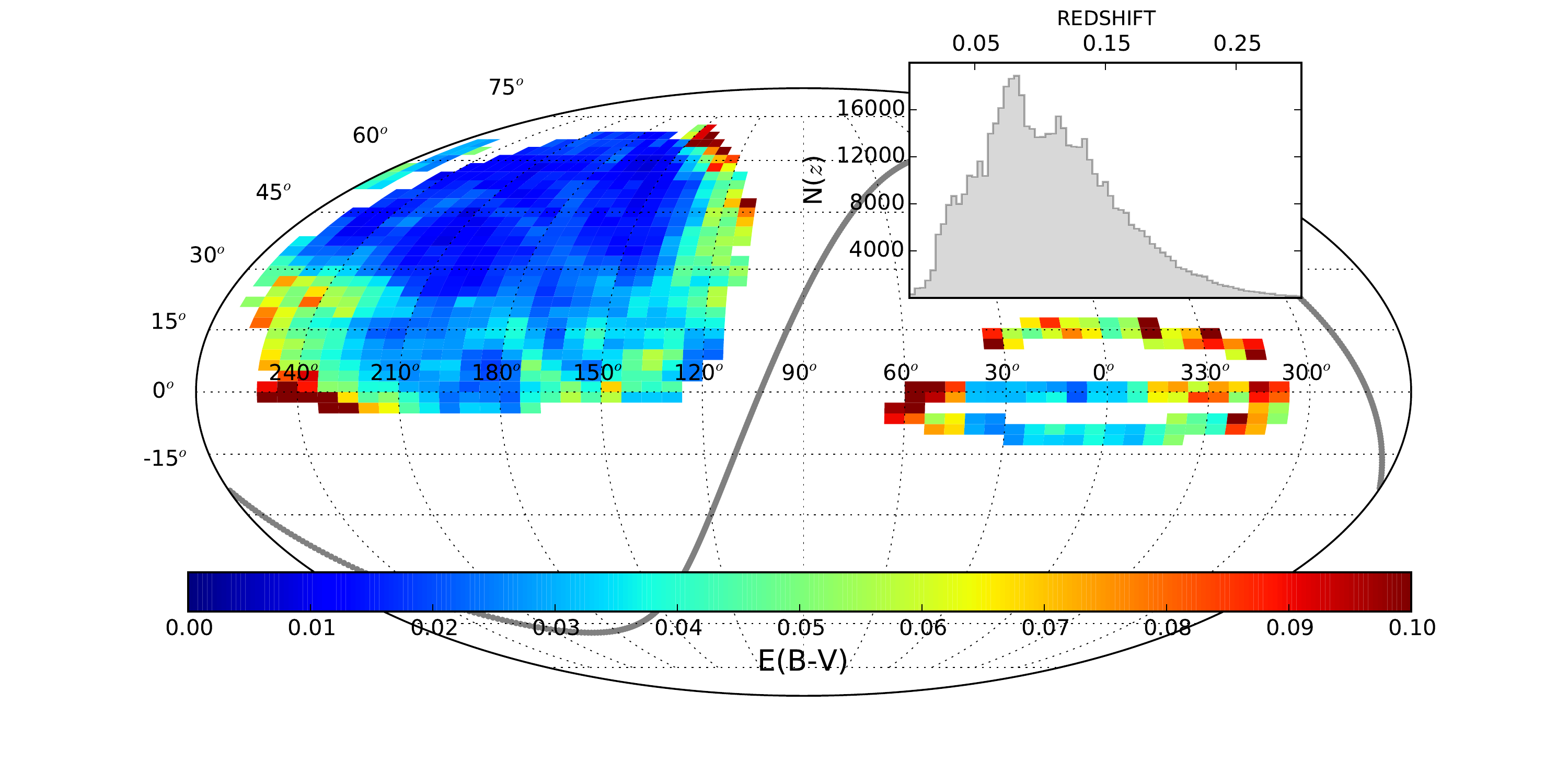}
\caption{Spatial distribution of the final sample of SDSS galaxies, with its spectroscopic redshift distribution (top-right inset). The color code indicates the average galactic extinction per pixel. The gray line shows the galactic plane.}
\label{map_ebv}
\end{figure}

The SDSS is a multi-band imaging and spectroscopic redshift survey using a dedicated 2.5-meter telescope at Apache Point Observatory in New Mexico. It provides deep photometry ($r<22.5$) in  $ugriz$ passbands. Our input data are selected from the data release 12 \citep[DR12,][]{Alam2015} by using the SDSS CasJob website (the MySQL query is given in Appendix A). From the SDSS database, we retrieved 516,525 sources classified as galaxy, with dereddened petrosian magnitudes $r\le 17.8$ and spectroscopic redshifts $z\le 0.4$. For all we queried the equatorial coordinates (RA, Dec), dereddened Petrosian magnitudes, ellipticities ($b/a$), galactic extinction \citep[]{Schlegel1998}, PSF Full-Widths-at-Half-Maximum (FWHMs) and sky background values in all bands. The spatial distribution of the final sample and its redshift distribution are shown in Figure~\ref{map_ebv}. The color code indicates the mean galactic reddening excess in each cell, which increases sharply in the vicinity of the galactic plane. 

We also retrieved the photometric redshifts of \citet[][hereafter B16]{Beck2016}, which are the only such redshifts available for comparison in DR12. They were computed with a k-NN method \citep[][local linear regression]{Csabai2007} that included five dimensions ($r$ magnitude and  $u-g$, $g-r$, $r-i$, $i-z$ colors) and trained with deep and high redshift spectroscopic surveys in addition to the SDSS. These photometric redshifts have similar or better accuracies than those inferred from random forests or prediction trees on the MGS sample \citep[][]{CarrascoKind2013, Carliles2010}, and may thus serve as reference for machine learning photometric redshifts based on photometric measurements.

\subsection{Image data}
\label{sec:cutout}
We downloaded all the ``corrected frames'' image data of the SDSS data release 8 from the SDSS Science Archive Server, including the 118 runs on Stripe 82 that are part of the release \citep{2011ApJS..193...29A}. The image headers include the astrometric fix applied to the SDSS data release 9 \citep{2011ApJS..195...26A}. The frames come background-subtracted and photometrically calibrated with an identical magnitude zero-point \citep{2008ApJ...674.1217P,2011AJ....142...31B}. All 4,689,180 frames and their celestial coordinates are indexed in a table for query with a homemade query tool\footnote{\url{http://github.com/ebertin/selfserver}}. The purpose of this tool is to provide the list of all images likely to overlap a given sky area.

For every entry in the galaxy sample we query the list of overlapping frames for each of the 5 SDSS filters and run the \textsc{SWarp} tool\footnote{\url{http://astromatic.net/software/swarp}} \citep{2002ASPC..281..228B} to resample to a common pixel grid and stack all the available image data. We rely on the WCS parameters in the input image headers \citep{2002A&A...395.1077C} for the astrometry. The result is a $64\times 64 \times 5$ pixel datacube in a gnomonic projection centered on the galaxy coordinates, and aligned with the local celestial coordinate system. The output pixel scale is identical to that of the input images (0.396 arcsec). We choose a L\'anczos-3 resampling kernel \citep{1979JApMe..18.1016D} as a compromise between preserving image quality and minimizing the spread of possible image artifacts. Other \textsc{SWarp} settings are also set to default, except for background subtraction, which is turned off.

The number of overlapping frames contributing to a given output pixel in a given filter ranges from 1 or 2 for ``regular" SDSS images, to 64 for some of the galaxies in Stripe 82.

No attempt is made to remove singular images, to mask out neighbors or to subtract light contamination from close bright sources, hence all galaxy images are included in the final dataset unmodified.

\begin{figure}
\includegraphics[width=1\columnwidth]{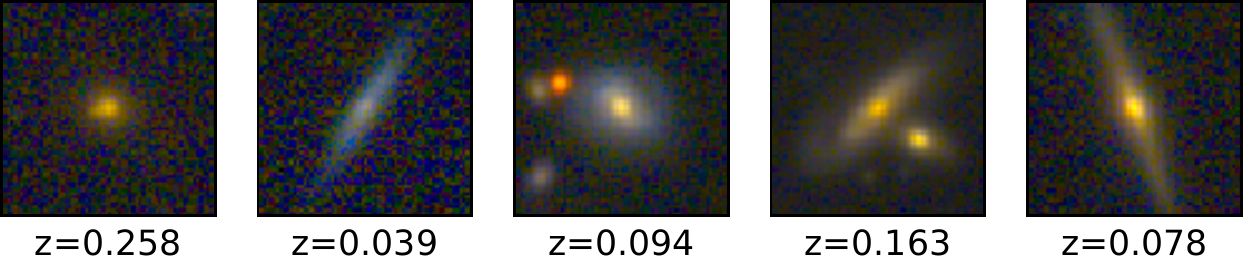}
\caption{Random examples from the image dataset. Images in the 5 SDSS channels were linearly remapped to Red,Green,Blue, with saturation $\alpha=2.0$ and display $\gamma=2.2$ applied following the prescriptions from \cite{2012ASPC..461..263B}.}
\label{fig:cutouts}
\end{figure}

Machine learning algorithms dealing with pixel data are generally trained with gamma-compressed images (e.g., in JPEG or PNG format). Yet our tests did not show any improvement of the classifier performance after applying dynamic range compression. We therefore decided to leave the dynamic range of the images intact.
Fig. \ref{fig:cutouts} shows a few cutouts from the dataset.

\begin{figure*}
\begin{center}
\includegraphics[width=0.9\textwidth]{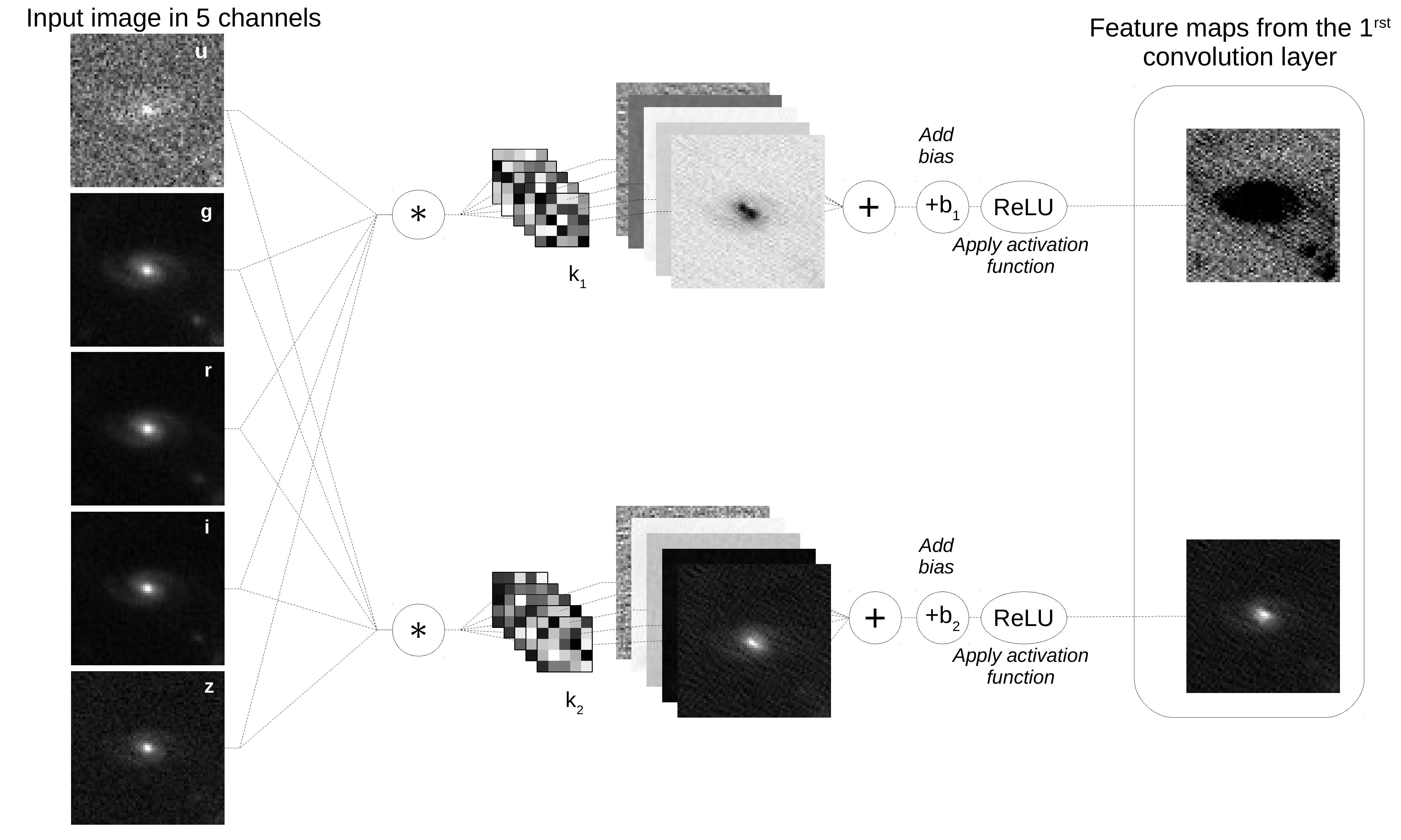}
\caption{Representation of the first convolution layer with two kernels of convolution to simplify the schema. The multi-spectral image of a galaxy in the five SDSS filters is fed to the network.
Each kernel is convolved with the multi-channel images. The convolved images are summed, an additive bias is added and a non-linear function is applied, to produce one feature map. } 
\label{conv}
\end{center}
\end{figure*}


\section{The Convolutional Neural Network} \label{sec:floats}
\subsection{Neural networks}
Artificial Neural Networks (NNs) are made of interconnected, elementary computing units called neurons. As in biology where a natural neuron is an electrically excitable cell that processes and transmits information via synapses connected to other cells, an artificial neuron receives a vector of inputs, each with a different weight, and computes a scalar value sent to other neurons. 
Once the type of neuron and the network topology have been defined, the network behavior relies solely on the strength of the connections (synaptic weights) between neurons. The purpose of learning (or training) is to iteratively adjust the synaptic weights until the system behaves as intended.

Multilayered neural network models are currently the basis of the most powerful  machine learning algorithms. In this type of NN, neurons are organized in layers. Neurons from a layer are connected to all or some of the neurons from the previous layer.

Convolutional Neural Networks (CNNs) are a special type of multilayered NNs. CNN classifiers are typically composed of a number of convolutional and pooling layers followed by fully connected layers.

\subsection{Convolutional layers}
A convolutional layer operates on a data cube, and computes one or several feature maps, also stored as a datacube. For the first convolution layer the input datacube is typically a multispectral image ($64\times 64 \times 5$ pixels in the case of our images taken through the $ugriz$ SDSS filters). Subsequent layers operate on feature maps from the previous layers.

The feature maps inside a convolutional layer are generally computed independently of one another. In our CNN, any given feature map relies on a set of adjustable convolution kernels, each of which is applied to the corresponding input data plane separately (with a stride of 1). The convolved arrays plus an adjustable offset are then summed and an \textit{activation function} applied element-wise to form the feature map.

Non-linear activation functions introduce non-linearity into the network. The most commonly used activation function is the ReLU \citep[Rectified Linear Unit,][]{ReLU} defined by $f(x)=\max(x,0)$. 
More recently, \cite{PReLU} introduced a Parametric ReLU (PReLU) with parameter $\alpha$ adjusted through basic gradient descent:

\begin{equation}
f(x) = \begin{cases} 
x & \textrm{if x  $\geq$ 0} \\ 
\alpha x & \textrm{if x $\leq $ 0}.
\end{cases}
\end{equation}

\subsection{Pooling layers}
Pooling layers reduce the size of input feature maps through down-sampling along the spatial dimensions ($2\times 2$ elements in our case). The down-sampling operation typically consists in taking the mean or the max of the elements.
Pooling allows feature maps to be computed on multiple scales at a reduced computational cost while providing some degree of shift invariance.

\subsection{Fully connected layers}
Contrary to their convolution layer equivalents, the neurons of a fully connected layer are connected to every neuron of the previous layer, and do not share synaptic weights.
In conventional CNNs, the fully connected layers are in charge of further processing the features extracted by the convolutional layers upstream, for classification or regression.

\subsection{Output layer}
\label{subsec:output}
We chose to handle the estimation of photometric redshifts as a classification problem, as opposed to using (non-linear) regression.
Previous studies \citep[e.g.,][]{Pasquet2017} demonstrated the benefits of this approach.
Each class corresponds to a narrow redshift bin $\delta z$. The galaxy at the center of the input multispectral image belongs to a single class (i.e. is at a single redshift).

The classes being mutually exclusive, we impose that the sum of the outputs be 1 by applying the \textit{softmax} activation function \citep{Bridle1990} to the output layer. The network is trained with the cross-entropy loss function \citep{Baum1987,Solla1988}. The output of this type of classifier was shown, both theoretically and experimentally, to provide good estimates of the posterior probability of classes in the input space \citep{Richard1991,Rojas1996}. This of course requires the neural network to be properly trained and to have enough complexity. 

\subsection{Our CNN architecture}
\label{architecture}
\begin{figure}
\includegraphics[height=0.94\textheight]{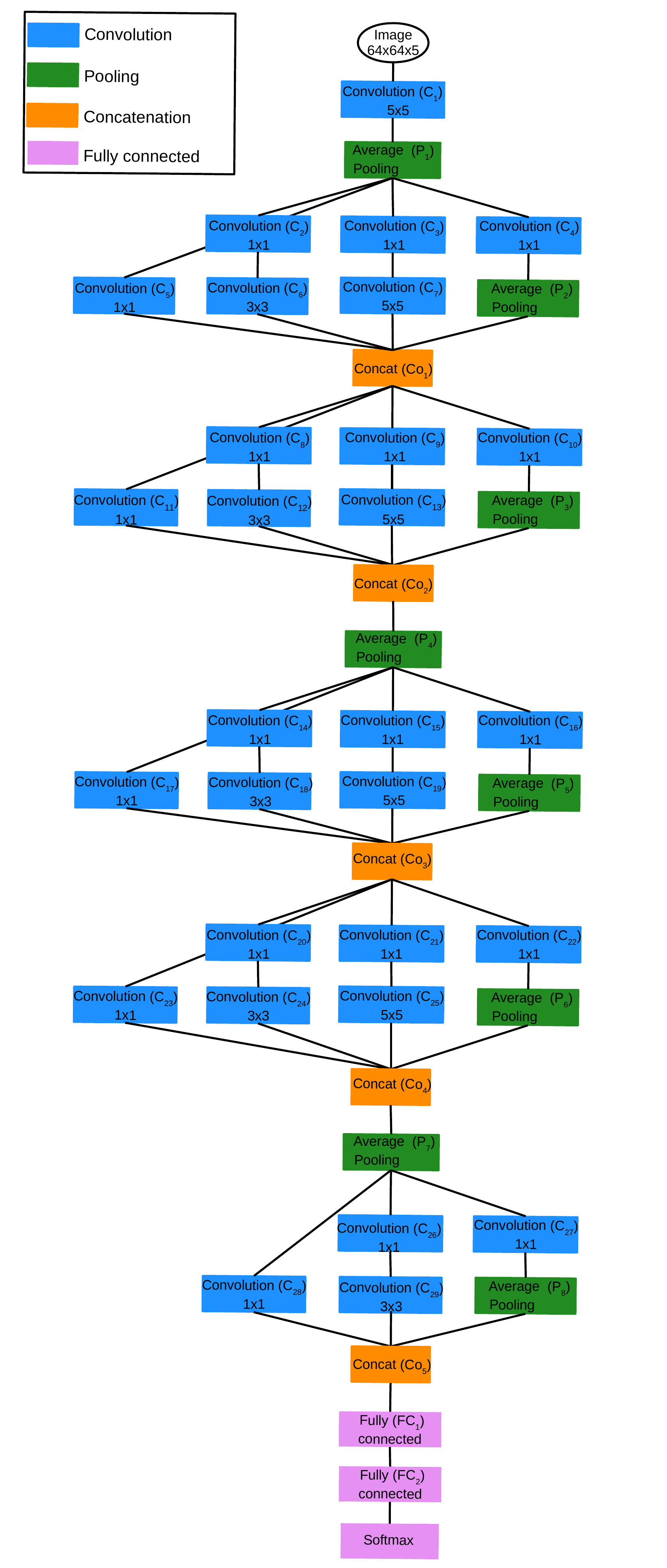}
\caption{Classifier architecture (see Section \ref{architecture} for a detailed description). The neural network is composed of a first convolution layer, a pooling layer and five inception blocks followed by two fully connected layers and a softmax layer. Further details can be found Table \ref{table_reseau}.}
\label{reseau}
\end{figure}

The overall architecture\footnote{The neural network model and examples are available at: \url{https://github.com/jpasquet/photoz}} is represented in Figure \ref{reseau}. The network takes as input a batch of images of size $64 \times 64$, centered on the galaxy coordinates, in five channels corresponding to the five bands $(u, g, r, i, z)$. 
The first layer performs a convolution with a large kernel of size $5 \times 5$. The pooling layer that follows reduces the size of the image by a factor of $2$. All the poolings in the network are computed by selecting the mean value in the sliding window. Although many studies use max pooling for classification problems, our experience with SDSS images is that average pooling performs better. The fact that most image pixels are dominated by background noise may explain why max pooling does not work as reliably.

We also found the PReLU activation function to perform better than the traditional ReLU in convolutional layers. One possible explanation may be that the negative part of the PReLU does not saturate, allowing faint signals below the threshold (e.g., those dominated by background noise) to propagate throughout the layer.

The remaining convolution part of the network is organized in multi-scale blocks called inception modules \citep{googlenet}. Each inception module is organized in two stages. In the first stage, the feature maps are convolved by three ``$1\times1$'' convolution layers. These layers are used to combine input feature maps and reduce their number before the more computationally expensive convolutions of the second stage. In this second stage, feature maps are processed in parallel in a pooling layer and a pair of larger convolution layers with size $3\times3$ and $5\times5$ to help identify larger patterns. Note that the last inception module does not include the $5\times 5$ convolution layer as the last feature maps ($8 \times 8$) have become too small to generate a useful output.

All the feature maps coming out from the last multi-scale block are concatenated and sent to a fully connected layer of 1096 neurons. One extra input containing the Galactic reddening value for the current galaxy is also added at this stage (Section \ref{reddening}).

After a second fully connected layer of 1096 neurons comes the output softmax layer. We set the number of classes to $N_c=180$ redshift bins over the range 0 - 0.4, with constant width $\delta z = 2.2\times 10^{-3}$. We believe that this sampling offers a reasonable compromise between the number of galaxies in each bin and redshift quantization noise. Such a large number of classes is not uncommon in modern classification challenges \citep[e.g.,][]{ImagenetChallenge}. As we shall see, the resulting distribution (a vector of 180 posterior probabilities) provides a reliable estimate of the photometric redshift PDF. 

The number and sizes of the feature maps are provided in Table \ref{table_reseau}. Appendix \ref{archi} also details the computational time needed for the various training steps.  

In total the number of adjustable parameters is 27,391,464. A common concern with large neural networks is overfitting, which happens whenever a model with too many free parameters learns irrelevant and noisy details of the training data to the extent that it negatively impacts its performance.
To make sure that our classifier was not overfitting, we monitored the loss on both the training and validation sets during the training stage. The PIT distribution (Section \ref{subsec:pdf}) also proved a valuable monitoring tool. Other methods we tested for addressing overfitting are batch normalization \citep{IoffeS15} and dropouts \citep{dropout}. However they did not improve the performance of our classifier.

\subsection{Photometric redshift estimation}
Although photometric redshift PDFs may be directly incorporated into larger Bayesian schemes for inferring model parameters (e.g., cosmological parameters), they may also be used to compute point estimates of the individual redshifts. We estimate the photometric redshift of a given galaxy by computing its expected value from its PDF $P(z_k)$: 
\begin{equation}
z_{phot}=\sum_{k=1}^{N_c} z_k\,P(z_k)\,,
\label{eq:newzphot}
\end{equation}
where $z_k$ is the midpoint value of the $k$th redshift bin. Assuming that $P(z_k)$ is reliable over the whole redshift range (see section~\ref{subsec:pdf}),   Eq.~\ref{eq:newzphot} provides a minimum mean square error estimation of the photometric redshift given the data.

Alternatively, we tested non-linear regression models with a multilayered neural network to estimate photometric redshifts directly from the softmax layer PDF output, using both a quadratic and Median Absolute Deviation (MAD) cost functions (see Section \ref{subsec:metrics}). 
After training with the spectroscopic redshifts, both models performed almost identically and did not provide any significant accuracy improvement over Eq.~\ref{eq:newzphot}.

\subsection{Experimental protocol}
\label{subsec:protocol}

\begin{figure}
\centering
\includegraphics[width=9cm]{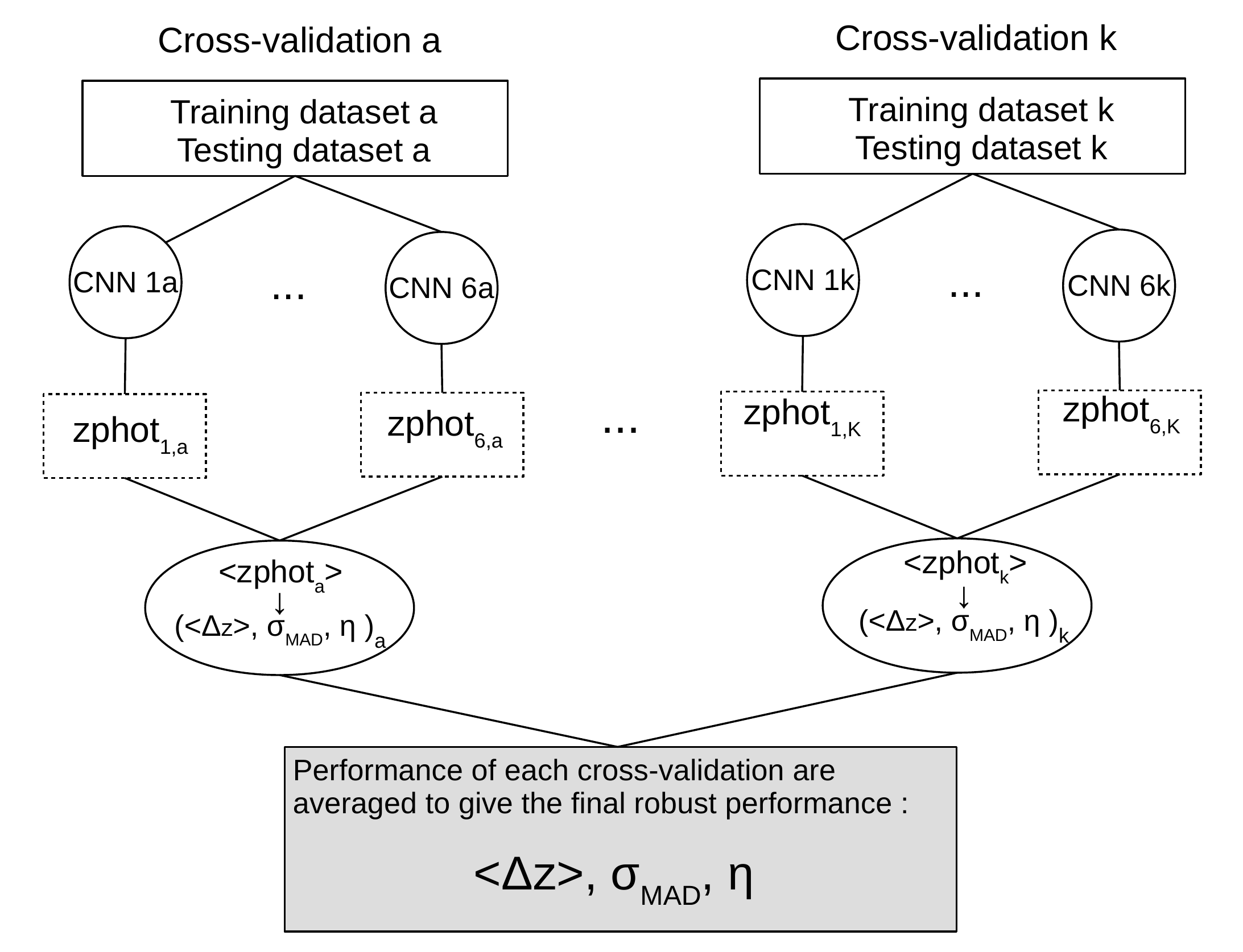}
\caption{Schema of the experimental protocol. For each training and testing datasets, six models are trained and averaged to get the final photometric redshift values. Moreover five cross-validation are performed. Their respective performances are averaged to provide the final scores (bias, $\sigma_{\rm MAD}$ and $\eta$).}
\label{proto}
\end{figure}

We divided the database into a training sample containing 80\% of the images and a testing sample composed of the remaining 20\%. 
To ensure that the CNN is not affected by galaxy orientation, we augmented the training set with randomly flipped and rotated (with $90\deg$ steps) images. 
We also selected 20,000 images in the training database to create a validation sample that allows us to control the performance of the model.

To increase the performance, we trained an ensemble of classifiers as it was shown to be more accurate than individual classifiers \citep[e.g.][]{NIPS2012_4824}. Moreover the generalization ability of an ensemble is usually stronger than that of base learners. 
This step involves training $N$ times one model with the same training database but a different initialization of the weights. As a compromise between computation time and accuracy, we chose $N=6$.
The individual decisions were then averaged out 
to obtain the final values of the photometric redshifts. 
We also averaged the PDFs, although we are aware that the result is a pessimistic estimate of the true probability distribution function. Other combination methods will be investigated in a future analysis. In the following sections, the terms photometric redshift and PDF refer to averages over the 6 trainings.
We also processed five cross-validations of the database to evaluate the stability of the network. This involves performing five learning phases, each with its own training sample (80\% of the initial database) and testing sample (the remaining 20\%), so as to test each galaxy once. 
 
Table \ref{tab:CNN stability} shows the performance of each cross-validation, using the metrics (bias, $\sigma_{\rm MAD}$ and $\eta$) defined in Section \ref{subsec:metrics}.
The bias varies slightly but the standard deviation and the fraction of outliers do not change significantly. Therefore we find the network to be statistically robust. In the following, the quoted values for the bias, standard deviation and fraction of outliers are the average values over 5 cross-validations, unless otherwise stated. 
Figure  \ref{proto} summaries the protocol of the training process.

\begin{table}
\begin{center}
\begin{tabular}{|c|c|c|c|}
\hline 
Cross validation & bias & $\sigma_{\rm MAD}$  & $\eta$    \tabularnewline
\hline 
1 & 0.00008 & 0.00914 & 0.31 \tabularnewline
\hline 
2 & 0.00009 & 0.00908 & 0.31 \tabularnewline
\hline 
3 & 0.00018 & 0.00913 & 0.32 \tabularnewline
\hline 
4 & 0.00011 & 0.00912 & 0.31 \tabularnewline
\hline 
5 & 0.00002 & 0.00910 & 0.29 \tabularnewline
\hline 
\textbf{mean} & 0.00010 & 0.00912 & 0.31 \tabularnewline
\hline 
\footnotesize{\textbf{standard deviation}}  & $5\times 10^{-5}$  & $2\times 10^{-5}$ & $9\times 10^{-3}$ \tabularnewline
\hline 
\end{tabular}\caption{Stability of the CNN over 5 cross-validations. The metrics (bias, $\sigma_{\rm MAD}$ and $\eta$) are defined in Section \ref{subsec:metrics}. The performance shows very low dispersion.}
\label{tab:CNN stability}
\end{center}
\end{table}


\section{Results}
In this section, we present the overall performance of our method for the estimation of photometric redshifts, and test the statistical reliability of the PDFs. 

\begin{figure}
\includegraphics[width=9.3cm]{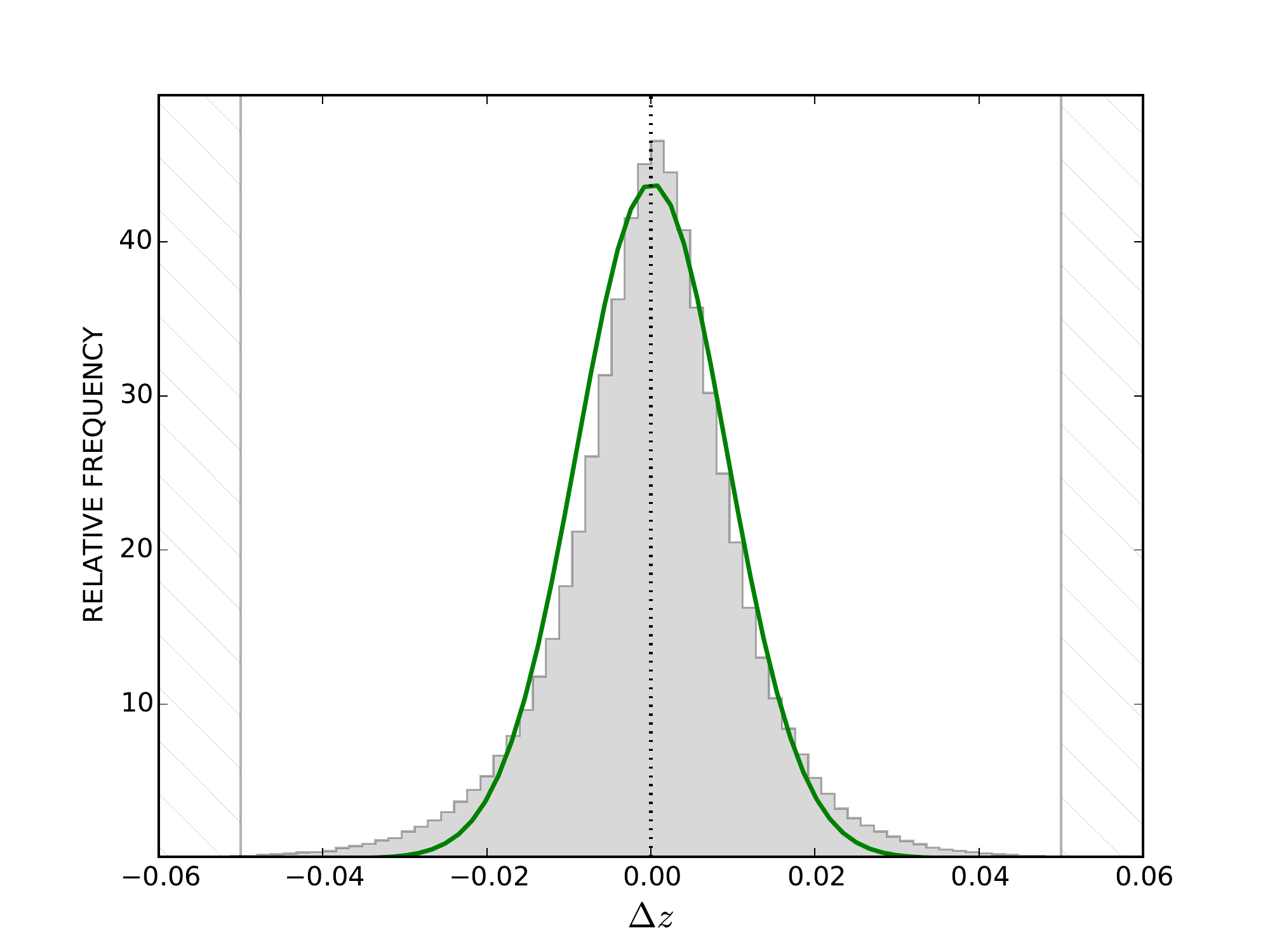}
\caption{Normalized $\Delta z$ distribution. The green line is a Gaussian distribution with the sigma and bias defined in Section \ref{subsec:metrics} and reported in Table \ref{tab:table_erreur}. The hatched parts define the catastrophic outliers.}
\label{deltaz_gauss}
\end{figure}

\begin{figure*}
\includegraphics[width=\textwidth]{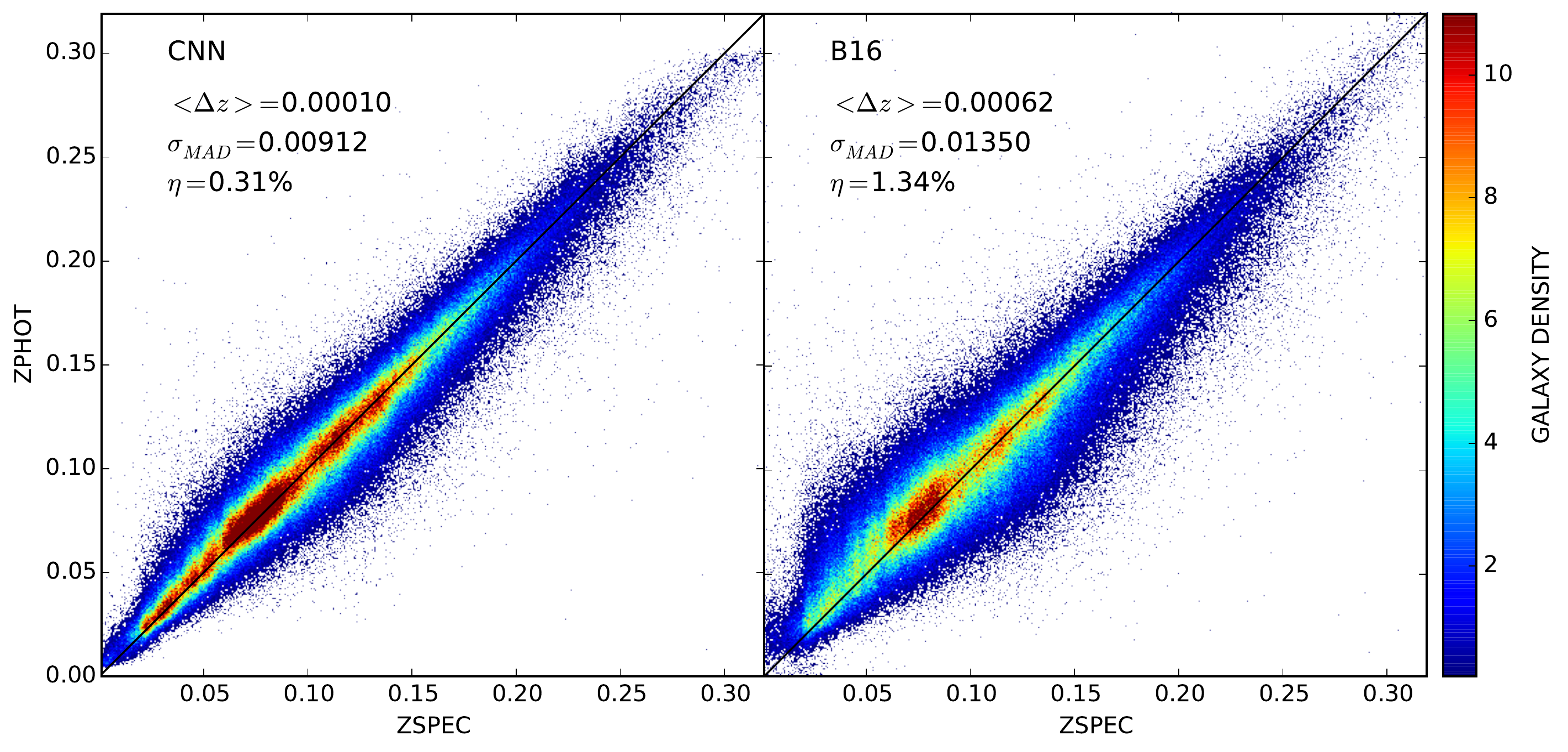}
\caption{Comparison between the photometric redshifts predicted by the CNN (left panel) and by B16 (right panel) against the spectroscopic redshifts. The galaxy density and the statistics are averaged over the 5 cross-validation samples.}
\label{histo_photoz}
\end{figure*}

\begin{figure}
\includegraphics[width=8.8cm]{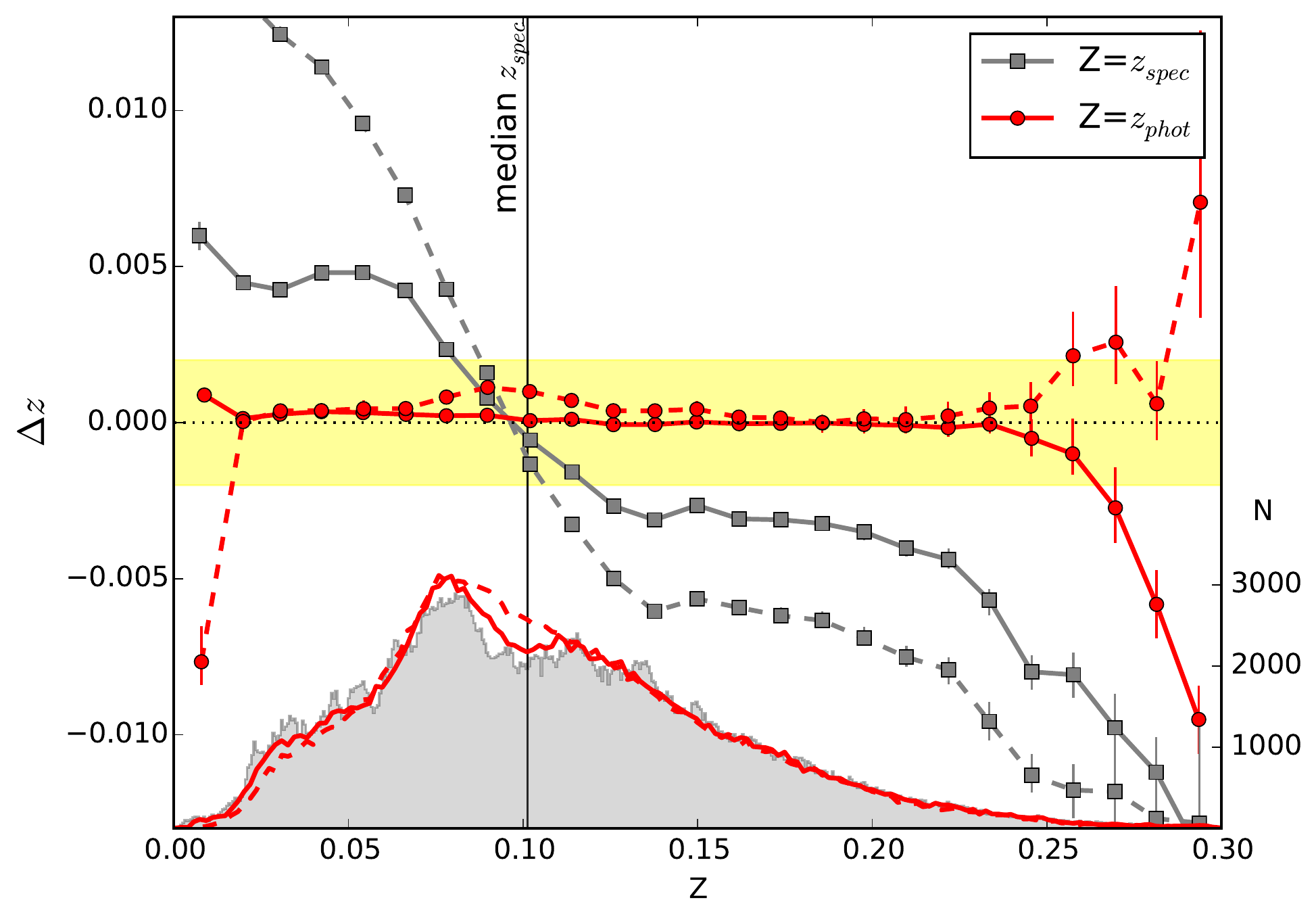}
\caption{Mean residuals as a function of spectroscopic redshifts (gray) and CNN redshifts (red). The dashed lines show the corresponding results for B16. The histograms are the respective redshift distributions. The CNN redshifts tend to be slightly over(under)-estimated below(above) the most populated redshift bins of the training sample, i.e. are biased towards the region of highest training. The effect is larger for B16, however no bias is found as a function of photometric redshift in either case.}
\label{z_bias}
\end{figure}

\subsection{Metrics}
\label{subsec:metrics}
 To assess the quality of the photometric redshifts,  we adopt the following commonly used statistics:
 \begin{itemize}
\item the \textbf{residuals},  $\Delta z= (z_{phot}-z_{spec})/(1+z_{spec})$, following the definition of \citet{Cohen2000};

\item the \textbf{prediction bias}, $<\Delta z>$, defined as the mean of the residuals;

\item the \textbf{MAD deviation} $\sigma_{\rm MAD}=1.4826\times {\rm MAD}$, where MAD (Median Absolute Deviation) is the median of $|\Delta z - \textrm{Median}(\Delta z)|$\footnote{Our definition of $\sigma_{\rm MAD}$ differs from the modified version adopted by \citet[]{Ilbert2006}, in which the median residual is not subtracted.}; 
\item the fraction  $\eta$ of \textbf{outliers} with $|\Delta z| > 0.05$, chosen to be $\sim$5 times the $\sigma_{\rm MAD}$ achieved by the CNN.
\end{itemize}

 The statistics for the network described in the previous section are reported in Table \ref{tab:CNN stability}, as well as in the first row of Table \ref{tab:table_erreur} (mean values over the 5 cross-validations). The normalized distribution of $\Delta z$ is shown in Figure \ref{deltaz_gauss}. The green line is a Gaussian distribution with the inferred sigma and bias. The hatched zones define the catastrophic outliers.

\begin{table*}[t]
\begin{supertabular}{|>{\raggedright}m{2cm}|>{\centering}p{3.5cm}|>{\centering}p{1.8cm}|>{\centering}p{1.8cm}|c|c|c|c|}
\hline 
\multicolumn{2}{|c|}{Trial} & {training sample size} & {size of 1 test sample} & {bias} & {$\sigma_{\rm MAD}$} & {$\eta$} & $<CRPS>$ \tabularnewline
\hline 
\multicolumn{2}{|>{\centering}p{6.3cm}|}{\bf Training with 80\% of the dataset} & 393,219  &   &  &   &   &  \tabularnewline
\multicolumn{2}{|>{\raggedleft}p{6.3cm}|} {Full test sample} &   & 103,306  & 0.00010  & 0.00912  & 0.31  & 0.00674 \tabularnewline
\multicolumn{2}{|>{\raggedleft}p{6.3cm}|} {(B16)} &  & (103,306) & (0.00062) & (0.01350) & (1.34) &  \tabularnewline
\multicolumn{2}{|>{\raggedleft}p{6.3cm}|} {Suspect zone (SZ) removed} &  & 101,499 & 0.00004 & 0.00908 & 0.31 & 0.00672\tabularnewline
\multicolumn{2}{|>{\raggedleft}p{6.3cm}|} {Widest 10\% of PDFs} &  & 91,543  &  0.00006 & 0.00848 & 0.09 & 0.00606 \tabularnewline
\multicolumn{2}{|>{\raggedleft}p{6.3cm}|} {Widest 20\% of PDFs} &  & 79,897 &  0.00005 & 0.00789 & 0.06 & 0.00556 \tabularnewline
\multicolumn{2}{|>{\raggedleft}p{6.3cm}|} {Stripe 82 only} &  & 3,943 &  -0.00009 & 0.00727 & 0.34 & 0.00574 \tabularnewline
\multicolumn{2}{|>{\raggedleft}p{6.3cm}|} {Stripe 82 with widest 20\% of PDFs removed} &   & 3,131 &  0.00004 & 0.00635 & 0.09 & 0.00467 \tabularnewline
  \hline 
\multicolumn{2}{|>{\centering}p{6.3cm}|}{Training with 50\% of the dataset$^\star$} & 250,000 & 252,500 & 0.00007 & 0.00910 & 0.29 & 0.00672 \tabularnewline
\hline 
\multicolumn{2}{|>{\centering}p{6.3cm}|}{Training with 20\% of the dataset} & 99,001 & 385,970 & -0.00001 & 0.00914 & 0.30 & 0.00677 \tabularnewline
\hline 
\multicolumn{2}{|>{\centering}p{6.3cm}|}{Training with 2\% of the dataset} & 10,100 & 434,228  &  -0.00017 & 0.01433 & 1.26 & 0.01009 \tabularnewline
\hline 
\multicolumn{2}{|>{\centering}p{6.3cm}|}{Training on Stripe 82} & 15,771 & &  &  &  &  \tabularnewline
\multicolumn{2}{|>{\raggedleft}p{6.3cm}|}{Stripe 82 removed$^\star$} & & 478,274 & 0.00194 & 0.01341 & 1.15 & 0.00988 \tabularnewline
\multicolumn{2}{|>{\raggedleft}p{6.3cm}|} {Stripe 82 only} &   &  3,942 &  -0.00002 & 0.00795 & 0.38  & 0.00622  \tabularnewline
\hline 
\multicolumn{2}{|>{\centering}p{6.3cm}|}{Training w/o Stripe 82} & 486,560 &  & & &  &  \tabularnewline
\multicolumn{2}{|>{\raggedleft}p{6.3cm}|}{Stripe 82 removed$^\star$} & & 97,607 & 0.00000 & 0.00914 & 0.33 & 0.00680 \tabularnewline
\multicolumn{2}{|>{\raggedleft}p{6.3cm}|} {Stripe 82 only$^\star$} &   &  19,714 & -0.00077 & 0.00760 & 0.41 & 0.00606 \tabularnewline
\hline 

\end{supertabular}
\caption{Statistics for various CNN trials, with the B16 results in parenthesis where the comparison is relevant. The bias, $\sigma_{\rm MAD}$ and fraction of outliers $\eta$ are defined in Section \ref{subsec:metrics}. The values are averaged over 5 test samples, except in the cases marked with a $\star$ where there is only one. The CRPS and PDF width are defined in section \ref{subsec:pdf}. The ``suspect zone" (SZ) was identified as a small region of the SDSS with above average bias (see Section \ref{subsec:suspect}). It has been removed by default in all other cases below the ``SZ removed" line.
}
\label{tab:table_erreur}
\end{table*}

\subsection{Photometric redshifts}
\label{subsec:zphot}
The distribution of the photometric vs spectroscopic redshifts obtained from the CNN (Fig. \ref{histo_photoz}, left panel) shows a striking decrease of the dispersion around the truth value compared to B16 (right panel), the latest and only comparable study. This is reflected in the scores for the three statistics: a factor of 1.5, 6 and 4 improvement for the $\sigma_{\rm MAD}$, bias and outlier rate respectively (the B16 values are listed in parenthesis in Table \ref{tab:table_erreur}). However we observe a plateau near z$\sim$0.3 for the CNN, where high redshift objects are significantly under-represented (galaxies with z$\sim$0.3 represent 0.1\% of the training database)\footnote{A preliminary test showed that the size of the training batches of images, progressively fed to the CNN, may be at least partly responsible for this plateau: doubling the size improved the prediction at the highest redshifts and moved the plateau upward. Better balancing the redshifts inside each batch may be necessary. This point will be addressed in a future analysis.}.
This trend is not observed in B16 as they use a larger training set that extend to much higher redshift.     

Figure \ref{z_bias} shows the bias as a function of both spectroscopic redshift and photometric redshift, with the corresponding redshift distributions and the B16 results for comparison. The photometric redshifts predicted by the CNN tend to be slightly over(under)-estimated below(above) the median redshift of the training sample, i.e. biased towards the most highly populated redshift bins. However the bias remains small and never exceeds 1${\sigma}$. It is also  significantly smaller than the bias induced by the B16 method. Most importantly, none is found as a function of photometric redshifts. This is particularly noteworthy for future large missions. The photometric redshift requirement for Euclid is $\Delta z \le 0.002$ in photometric redfshift bins (yellow shaded zone), well achieved by the present method. 

To understand the origin of the spectroscopic redshift dependent bias, one must realize that neural networks are naturally sensitive to priors in the training sample. In the case of our classifier, the PDF should be a good estimation of the redshift posterior probability density, given an input datacube $\vec{x}$ (Section \ref{subsec:output}). In virtue of the Bayes rule, and in the continuous limit, the PDF writes:
\begin{equation}
\label{eq:pzxi}
p(z|\vec{x}) = \frac{p(\vec{x}|z) p(z)}{\int p(\vec{x}|z) p(z) dz},
 \end{equation}
where $p(z)$ is the normalized redshift distribution of input galaxies in the training set, and $p(\vec{x}|z)$ the normalized distribution of galaxies with true redshift $z$ in multispectral image space. The net effect of $p(z)$ on the estimated photometric redshift $z_{phot} = \mathbb{E}[z|\vec{x}]$ is to ``push'' $z_{phot}$ in the direction of increasing $p(z)$. The detailed analysis of redshift-dependent biases will be the subject of a future paper.

\begin{figure*}
\begin{center}
\includegraphics[height=23.5cm]{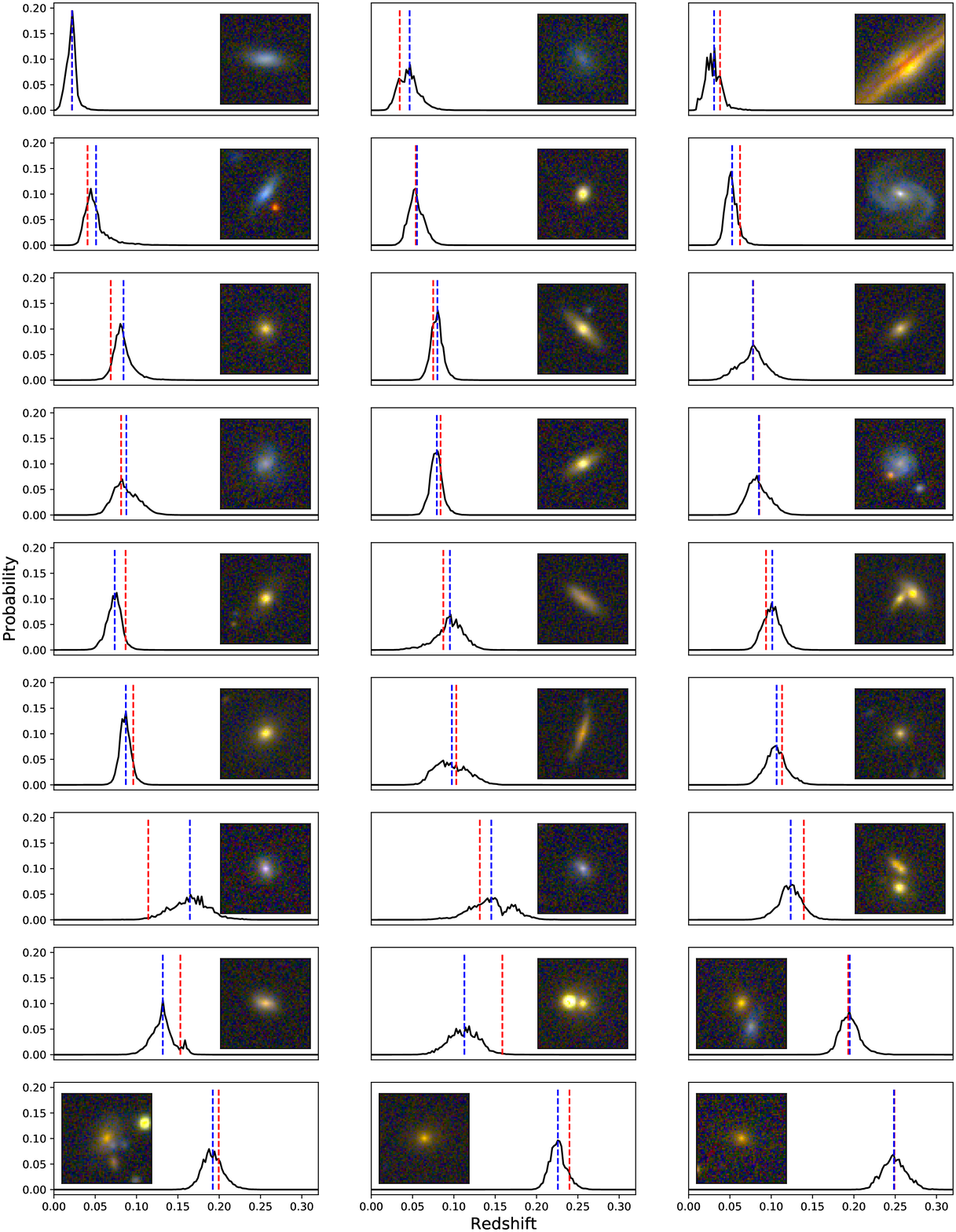}
\caption{Random examples of PDFs obtained from the CNN. The red dotted lines mark the values of the spectroscopic redshifts and the blue ones the photometric redshifts estimated by the CNN and computed as the softmax weighted sum of the redshift values in each bin of the PDF.}

\label{pdf}
\end{center}
\end{figure*}

\begin{figure}
\includegraphics[width=8.5cm]{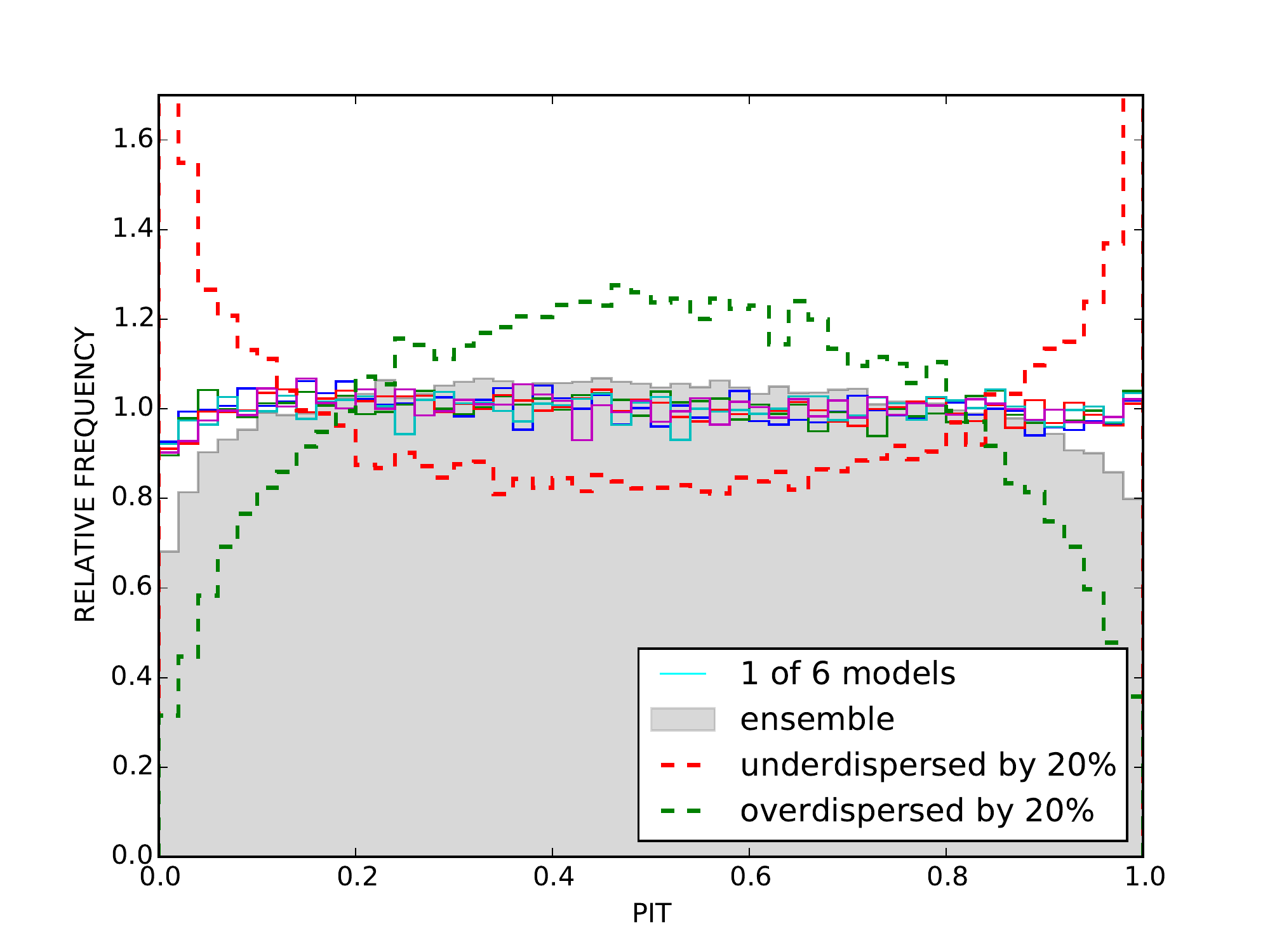}
\caption{Distribution of the Probability Integral Transforms (PIT) for each of the six models, and for the final PDFs. Each model exhibits a nearly flat PIT distribution, which assesses the very good quality of the PDFs. The PIT distribution of the final (averaged) PDFs (see Section \ref{subsec:protocol}) is slightly over dispersed, as expected from our pessimistic choice of combination. The dashed green and red lines result from expanding and shrinking, respectively, the $\Delta z$ of a flat model by 20\%.
}\label{fig:pit}
\end{figure}

\subsection{Probability Distribution Functions}
\label{subsec:pdf}
The PDFs of a subset of randomly selected galaxies from the test set are shown in Fig.~\ref{pdf}, together with the galaxy RGB images. As proposed by \citet[]{Polsterer2016}, we use two statistics to assess the overall prediction quality of our PDFs: the Probability Integral Transform (PIT) and the Continuous Ranked Probability Score (CRPS). 

The PIT statistic \citep{dawid.2307/2981683} is based on the histogram of the cumulative probabilities (CDF) at the true value, i.e. the spectroscopic redshift. For galaxy $i$ at spectroscopic redshift $z_i$, with redshift probability distribution function $PDF_i$, the PIT value is:
\begin{equation}
CDF_i(z_i) =\int_{0}^{z_i} PDF_i(z)dz.
\label{cdf}
\end{equation}

A flat PIT distribution indicates that the PDFs are not biased with respect to the spectroscopic redshifts and are neither too narrow nor too wide, whereas convex or concave distributions point to under or over-dispersed PDFs, respectively \citep{Polsterer2016}. Excessively narrow PDFs will often miss the target, overproducing PIT values close to 0 or 1, whereas PDFs that are too wide will encompass the true redshifts more often than not and therefore favor intermediate PIT values. 
The PIT distribution for each of the six models in our ensemble of classifiers, and for the final PDFs (see Section \ref{subsec:protocol}) are shown in Figure \ref{fig:pit}.  Each individual model exhibits a nearly flat PIT distribution, indicating well behaved probability distribution functions. The PIT distribution of the final (averaged) PDFs is slightly over dispersed, as expected from our pessimistic choice of combination. 

The CRPS is a performance score 
\citep[well known in meteorological predictions,][] {crps:10.1175/1520-0434(2000)015<0559:DOTCRP>2.0.CO;2} that quantifies how well the predicted PDF represents the true spectroscopic redshift.  
For galaxy \textit{i}, it is defined as:
\begin{equation}
CRPS_i = \int_{-\infty}^{z_i} CDF_i(z)^{2}dz + \int_{z_i}^{+\infty} (CDF_i(z)-1)^{2} dz.
\end{equation}
The mean CRPS ($\sim 0.007$) is reported in the last column of Table~\ref{tab:table_erreur}. This value is significantly lower than the CRPS quoted by \citet{disanto2018} or \citet{Tanaka2018} ($\sim$ 0.1 and 0.02 respectively), 
although a fair comparison is difficult as these studies encompass larger redshift domains. However, the small mean CRPS and the nearly flat PIT distribution reflect the high reliability of our PDFs. 
 
To further assess the quality of the PDFs, we measure their "widths", defined as the redshift interval resulting from chopping off their left and right wings in equal measure, so as to keep 68\% of the probability distribution\footnote{Similar results are found when defining the PDF width as containing 90\% of the  probability distribution instead of 68\%.}.
Removing the widest 10\% (20\%) of the PDFs (width $>0.0383$ (0.0335)) significantly improve $\sigma_{\rm MAD}$ and $\eta$,
as reported in Table \ref{tab:table_erreur}. These improvements reinforce our confidence that our PDFs properly reflect the photometric redshift uncertainties.

\section{Size of the training database}
As acquiring large spectroscopic samples for training is very observing time intensive, it is crucially important to assess the performance of our CNN as a function of training size. 
Our baseline network made use of 400,000 galaxies (80\% of the database). We trained the same model on 250,000 and 100,000 galaxies (50\% and 20\% of the database respectively), and also adapted the network by reducing its depth and width for a 10,000 galaxy training sample (2\% of the database). 

The statistics of these three trials are reported in Table \ref{tab:table_erreur}. We find practically no fall in performance between the training on 400,000 objects and that on 100,000, which is a particularly encouraging result. Although the global statistics deteriorate significantly with 10,000 training sources, they remain comparable to the results of B16, which is also remarkable. 
Moreover, for all 3 trials including the 10,000 sources training sample, all the trends, or lack thereof, plotted in the next section remain nearly indistinguishable from our baseline 400,000 sources training case.


\section{Further behavioral analysis of the CNN} 
\label{behavior}

In this section, we study how the performance of the CNN varies with
a number of characteristics of the input images relating to galaxy properties or observing conditions.

\subsection{Galactic reddening}
\label{reddening}
\begin{figure*}
\centering
\includegraphics[width=8.8cm]{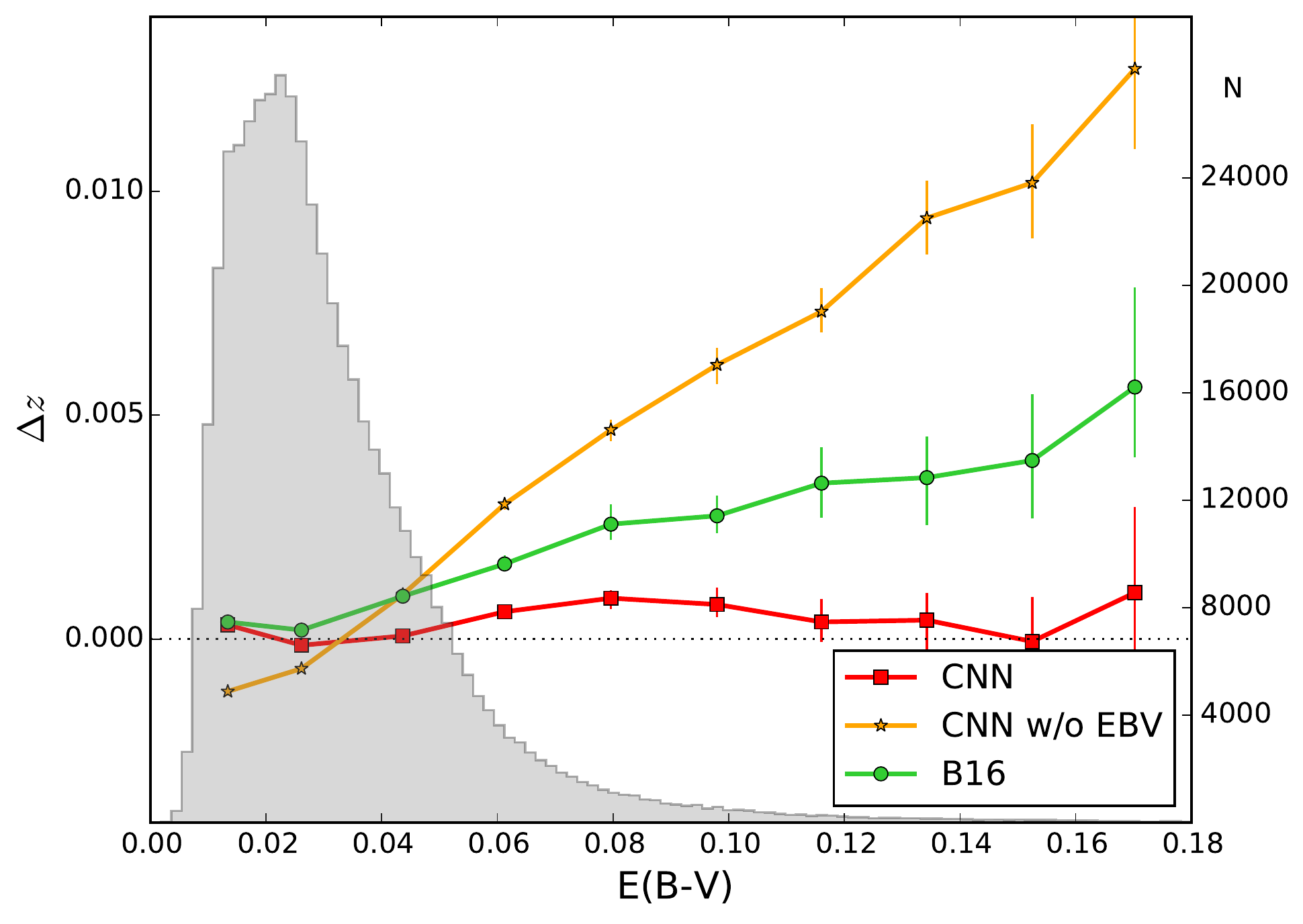}
\includegraphics[width=8.8cm]{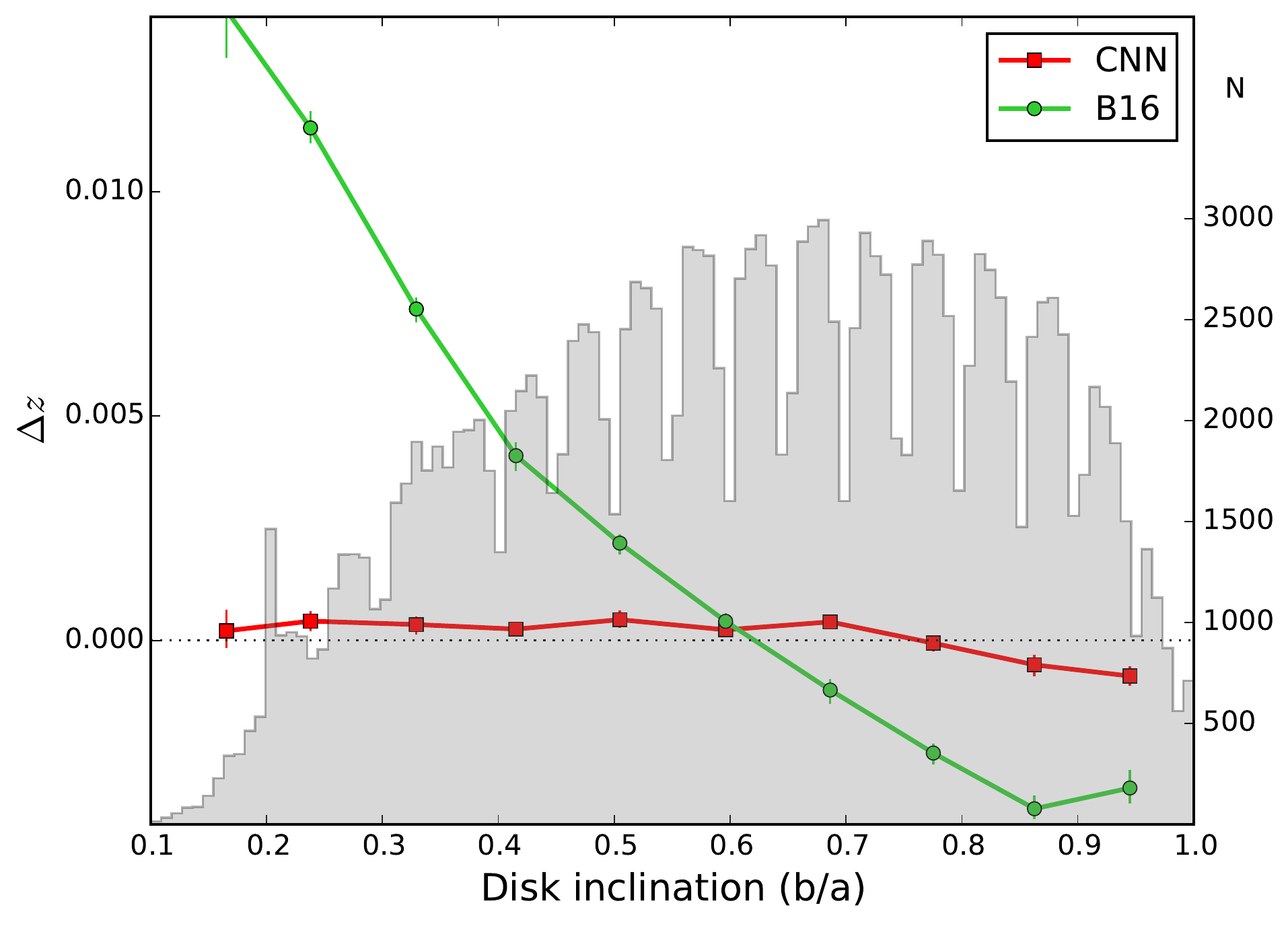}
\caption{\textbf{Left:} bias as a function of Galactic extinction for the classifier with and without integrating $E(B-V)$ into the training (red and orange lines respectively) and for B16 (green). The CNN tends to overestimate redshifts in obscured regions (confusing galactic dust attenuation with redshift dimming), unless  $E(B-V)$ is used for training.\textbf{ Right:} bias as a function of disk inclination for galaxies classified as star-forming or starburst. The CNN is virtually unbiased wrt ellipticity, unlike B16.}
\label{ebv}
\end{figure*}

As illustrated in Fig.~\ref{map_ebv}, the SDSS spans a large range in Galactic extinction. The effect of $E(B-V)$ on the observed colors of a galaxy can  mimic that of a redshift increase. The impact of including the reddening information into the training, or not, is illustrated on Fig.~\ref{ebv} (left panel). Without this information provided to the classifier, a strong reddening-dependent bias is observed (orange line). A weaker trend is observed for B16, who use SDSS de-reddened magnitudes. The small size of the sample at high galactic extinction most likely prevents the CNN from properly accounting for this degeneracy, hence our choice to include the reddening information into our model, which successfully removes the trend (red line). 

\subsection{Galaxy inclination}
As Galactic reddening, the inclination of a galaxy reddens its color but also affects the shape of the attenuation curve in a complex way \citep[]{Chevallard2013}. It has been a major issue for the estimation of photometric redshifts, especially with SED fitting codes \citep{Arnouts2013}. Figure~\ref{ebv} (right panel) shows that the CNN is very robust to galaxy inclination, unlike the B16 method, which is strongly biased, especially at high inclination. The network is able to account for this effect thanks to the large training sample, helped by the data augmentation process that further expanded it by rotating and flipping the images.
While B16 only uses the photometric information, \citet[][]{yip2011ApJ...730...54Y} showed that machine learning methods can better handle this bias if the inclination information is included into their training. 

\subsection{Neighboring galaxies}
Using the SDSS neighbors catalog, we investigated the performance of the CNN on crowded images (i.e. with at least one neighbor within 17 arcsec of the spectroscopic target). No statistical effect was found on the photometric redshift accuracy of the central targets. A few examples can be seen in Fig.~\ref{pdf}. 

\subsection{Variations throughout the surveyed area}
\label{subsec:suspect}

\begin{figure*}
\centering
\includegraphics[width=9.1cm]{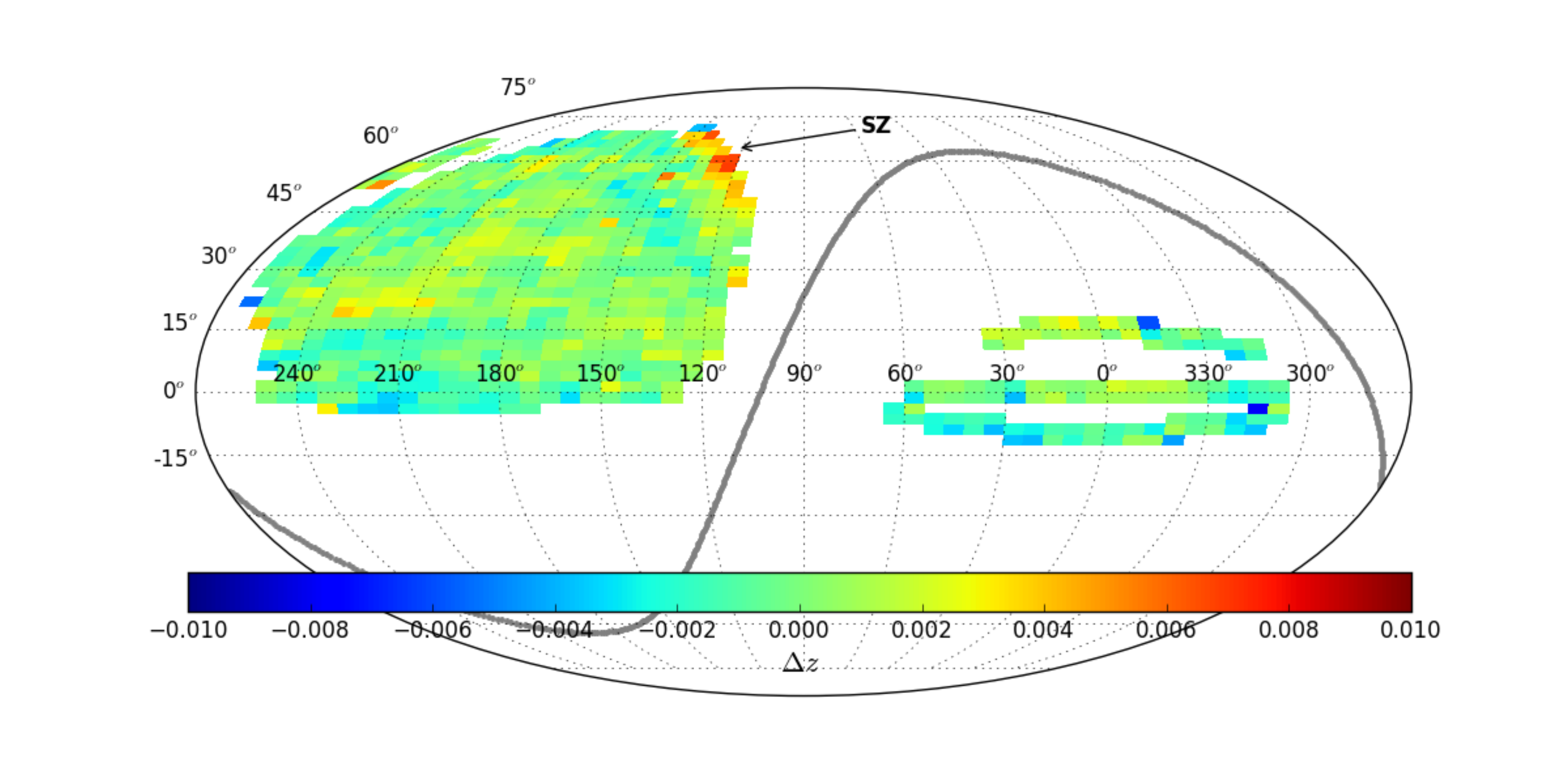}
\includegraphics[width=9.1cm]{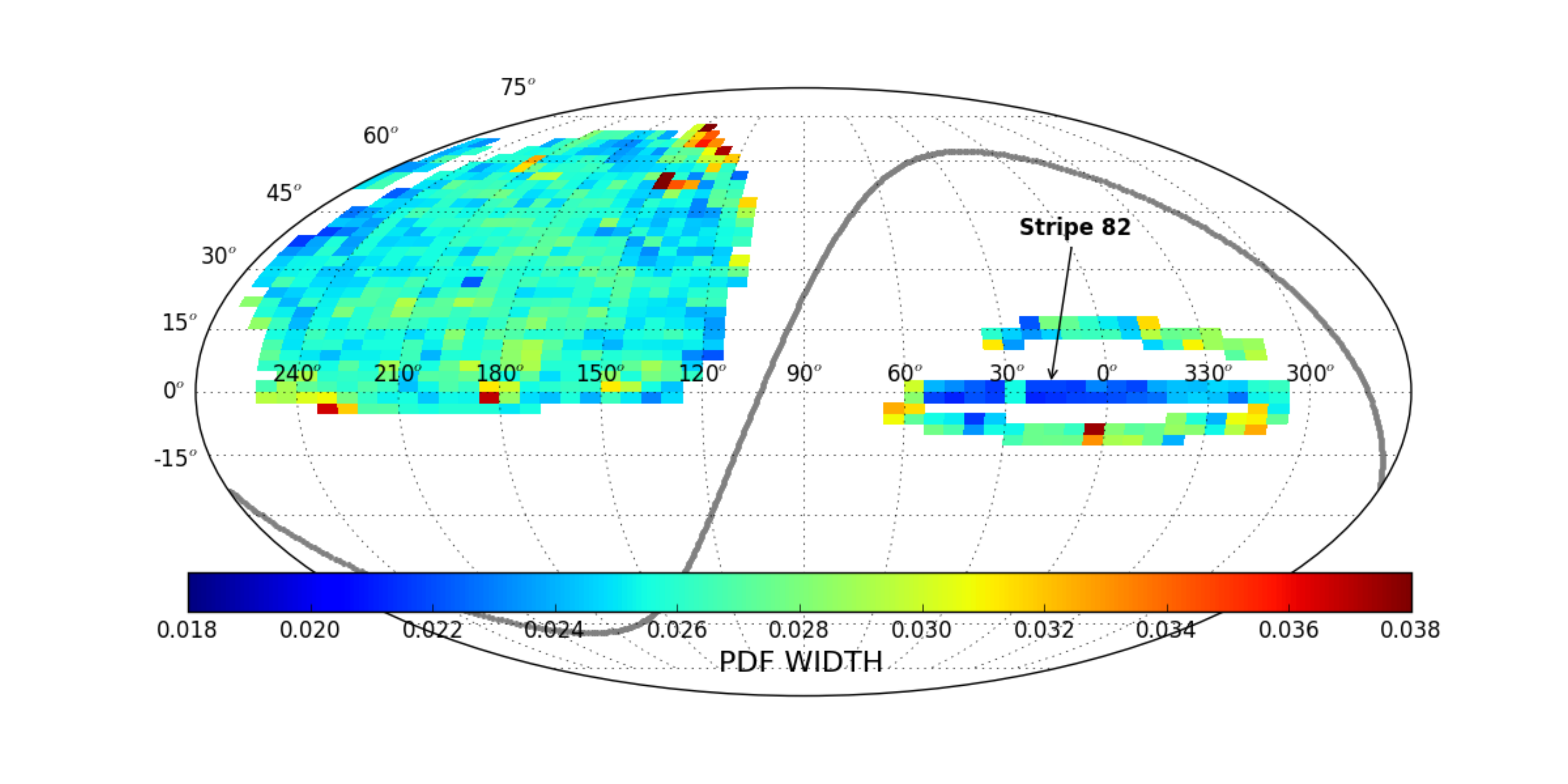}
\caption{Spatial distribution of the mean bias (left) and of the mean PDF width (right). The locations of the ``suspect zone" (SZ) and of the Stripe 82 region (see Section \ref{subsec:suspect}) are indicated.}
\label{map_quality}
\end{figure*}

Figure \ref{map_quality} shows the spatial variations of the bias and of the PDF widths on the celestial sphere (the color code refers to the mean quantities per cell). Overall, both quantities, which we find to be uncorrelated, show little variation throughout the surveyed area. However we identified a small region of the SDSS ($\sim 2.4\%$) where both are puzzlingly below average in quality (red patch towards the pole). 

This ``suspect zone" (SZ) appears to coincide with Stripe 38 and 39, but with no evidence of sky background, PSF or photometric calibration issues.  Although it is in a region of high Galactic extinction, excess reddening doesn't seem to cause the problem as it is not detected in the other regions of equally high galactic extinction (Fig.~\ref{map_ebv}). The bias on this patch alone is $\sim$30 times larger than on the full test sample (+0.0033 versus +0.0001), and also $\sim$10 times larger for B16. 
Removing the region from the test sample reduces the bias by a factor of 2.5, while $\sigma_{\rm MAD}$ is unaffected (see Table \ref{tab:table_erreur}).

\begin{figure}
\includegraphics[width=8.7cm]{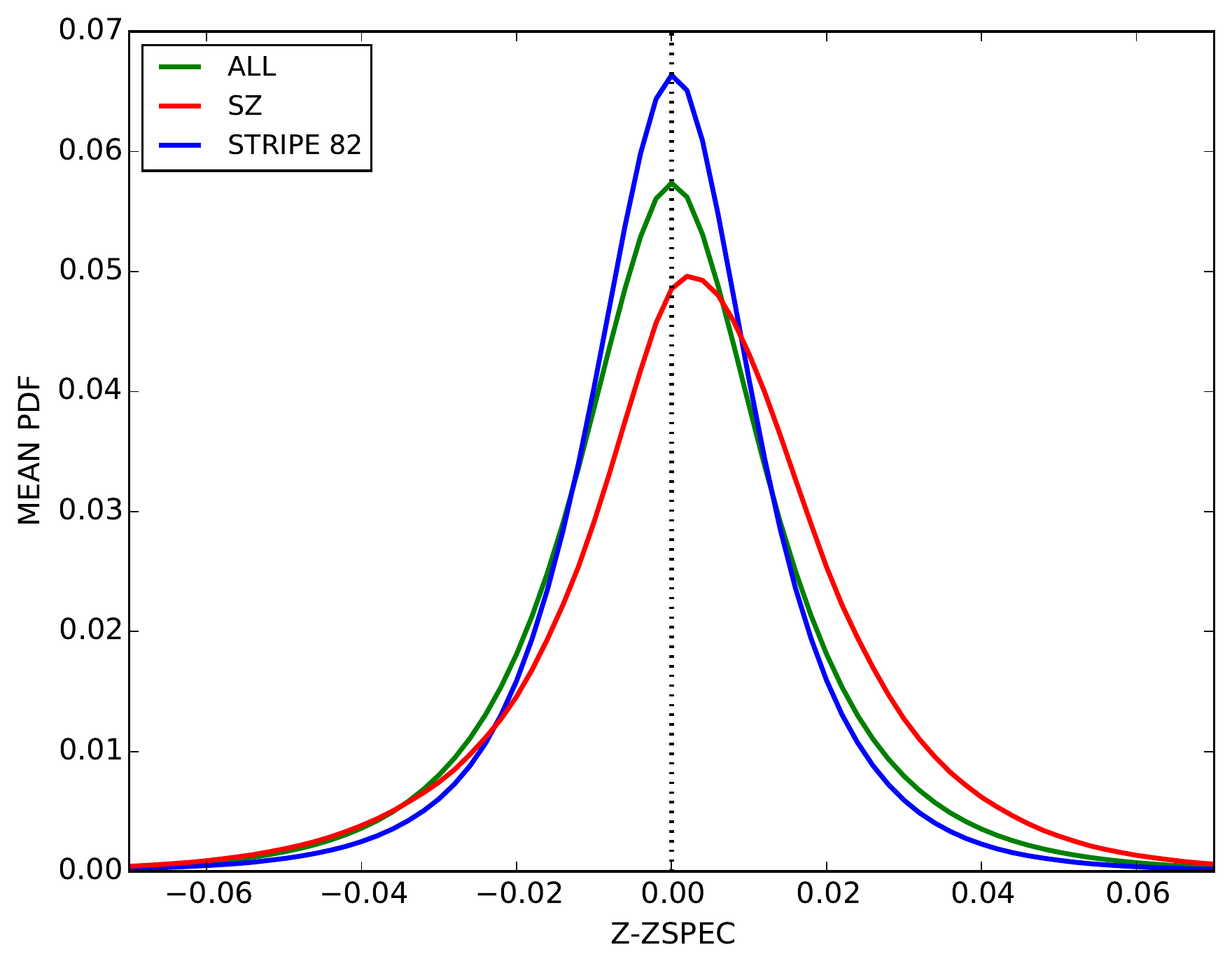}
\caption{The mean PDF in the SDSS patch showing defective photometric redshift and PDF quality (SZ) and in the Stripe 82 region, compared to the mean PDF of the full sample. The PDFs have been translated so that 0 corresponds to the spectroscopic redshifts.}
\label{3meanpdfs}	
\end{figure}

Also noteworthy is the Stripe 82 region, which exhibits narrower than average PDFs (dark blue stripe in the right panel of Fig.~\ref{map_quality}). This point is addressed in the next section. Figure \ref{3meanpdfs} shows the mean PDF in these two atypical regions compared to that of the full sample.

\subsection{Effect of noise}
\label{subsec:stripe82}
The Stripe 82 region, which combines repeated observations of the same part of the sky, gives us the opportunity to look into the impact of Signal-to-Noise Ratio (SNR) on our photometric redshift estimations. The statistics for this region alone are reported in Table \ref{tab:table_erreur}. The resulting $\sigma_{\rm MAD}$ outperforms that of the other tests (and can be further reduced by removing the widest PDFs). Thus increasing the SNR improves the performance of the classifier, even though the training was predominantly done using images with lower SNRs. 

We further tested the impact of SNR by training the same model (Fig.~\ref{reseau}) on two different datasets: one with Stripe 82 images only, one without Stripe 82. The statistics are reported in Table \ref{tab:table_erreur}. Removing Stripe 82 from the training set has no impact on the performance of the network outside of Stripe 82, unsurprisingly given the small perturbation it induces, and only slightly degrades $\sigma_{\rm MAD}$ on Stripe 82. This confirms that a CNN network mostly trained on low SNR images performs better on higher SNR images.

Evaluating whether training on high SNR images improves the performance on low and/or high SNR images is more difficult.
Training the network on Stripe 82 images reduces the training set to only $\sim 16,000$ galaxies, a very small sample for Deep Learning. The testing on Stripe 82 shows that $\sigma_{\rm MAD}$ is slightly higher and the bias lower, compared to training with the full dataset: a better match between the training and test sets may be compensating for the reduced training size.
The performance outside of Stripe 82 is degraded; this may be due to mismatched datasets (Stripe 82 images are too deep for the CNN to be able to classify shallow images well enough) and/or to the small size of the training sample. 
We can only conclude that small training sets with higher SNR images do not help the performance on low SNR images.

A more detailed analysis of this effect is presented in Fig.~\ref{performances appendix}, which shows the behavior of the redshift bias and $\sigma_{\rm MAD}$ as a function of SNR in all 5 bands. The SNRs were derived simply from the Petrosian magnitude errors quoted in the SDSS catalog. Stripe 82 galaxies were removed from this analysis, as the SDSS DR12 catalog does not provide photometric errors for the stacked images used in this work (see Section~\ref{sec:cutout}). As may be expected from the results on Stripe 82, $\sigma_{\rm MAD}$ gradually improves towards the highest SNR in all bands, with values lower than $\sim$0.006 for SNR$\ge$200 ($\sim$17,000 sources). 

$\sigma_{\rm MAD}$ seems to plateau at the low-end of SNRs, an effect that is not seen with B16. 
This is perhaps surprising but good news for the faintest sources, although it must be taken with a grain of salt as our SNR derived from Petrosian magnitude errors may be unreliable for faint objects.
The redshift bias shows no clear trend within the uncertainties  at low SNR but increases at high SNR. As high SNR objects are preferentially at low redshift ($z\le 0.1$), it probably simply reflects the bias discussed in Section \ref{subsec:zphot}, where galaxies with spectroscopic redshifts below the peak of the training set distribution have their photometric redshifts slightly over-estimated. 

\subsection{Influence of the PSF}

As Fig. \ref{performances appendix} shows, $\sigma_{\rm MAD}$ appears to be relatively immune to PSF variations, with only a slight increase for the worst observing conditions in individual bands (middle panel) or in case of large PSF variations between 2 bands (right panel).  
On the other hand, the redshift bias shows some trend with seeing (middle and right panels), similar to what is seen in B16, but with opposite signs. Larger PSFs generate an apparent decrease of the apparent surface brightness of galaxies that are not well resolved.
Note that SDSS observations are carried out through the different filters within a few minutes of interval, and therefore under very similar atmospheric conditions. This situation is likely to be worse for imaging surveys where the different channels are acquired during separate nights or even runs. Such datasets may require PSF information to be explicitly provided as input to the classifier in addition to the pixel and extinction data. 
 

\section{Summary and Discussion}
In this paper we have presented a Deep Convolutional Neural Network (CNN) used as a classifier, that we trained and tested on the Main Galaxy Sample of the SDSS at $z\le 0.4$, to estimate photometric redshifts and their associated PDFs. Our challenge was to exploit all the information present in the images without relying on pre-extracted image or spectral features. The input data consisted of 64$\times$64 pixel \textit{ugriz} images centered on the spectroscopic target coordinates, and the value of galactic reddening on the line-of-sight. We tested 4 sizes of training set: 400k, 250k, 100k and 10k galaxies (80\%, 50\%, 20\% and 2\% of the full database, respectively). 

In all cases but the last, we obtain a MAD dispersion $\sigma_{\rm MAD}=0.0091$. This value is significantly lower than the best one published so far, obtained from another machine learning technique (KNN) applied to photometric measurements by \citet[]{Beck2016} for the same galaxies ($\sigma_{\rm MAD}=0.0135$).
Restricting the training set to only 10,000 sources (although the CNN was not optimized for such a small number) increases dispersion by 60\%, but is still competitive with the current state-of-the-art.

The bias shows a quasi-monotonic trend with spectroscopic redshift, largely due to the prior 
imposed by the training set redshift distribution, as expected from PDFs behaving as posterior probabilities. However, the bias is independent of photometric redshift and lower than $10^{-4}$ at $z\le 0.25$, far below the $0.002$ value required for the scientific goals of the future Euclid mission.

We also find that: 1/ our photometric redshifts are essentially unbiased with respect to galactic extinction and galaxy inclination; 2/ the PDFs have very good predictive power, with a nearly flat distribution of the Probability Integral Transforms (PIT). Removing the widest PDFs improves the already small $\sigma_{\rm MAD}$ and fraction of outliers; 3/ $\sigma_{\rm MAD}$ decreases with the signal-to-noise ratio (SNR), achieving values below $0.007$ for SNR $>100$, as in the deep stacked region of Stripe 82; 4/ Variations of the PSF FWHM induce a small but measurable amount of systematics on the estimated redshifts, which prompts for the inclusion of PSF information into future versions of the classifier.

We conclude that, with a moderate training sample size ($\le$ 100,000), the CNN method is able to extract the relevant information present in the images to derive photometric redshifts and associated redshift PDFs whose accuracy surpasses the current state-of-the-art. 

The dependency of $\sigma_{\rm MAD}$ with SNR suggests that we have reached a point where the precision of individual photometric redshifts in the SDSS is essentially limited by image depth, not by the method.

This work opens very promising perspectives for the exploitation of large and deep photometric surveys, which encompass a larger redshift range and where spectroscopic follow-up is necessarily limited. New issues will arise regarding the representativity of the galaxy population in the spectroscopic samples across the whole redshift range, that will require dedicated investigations \citep[e.g.][] {Beck2017MNRAS.468.4323B} in anticipation of the LSST and Euclid surveys.




\vspace{1cm}
\section*{Acknowledgements}
This work has been carried out thanks to the support of the OCEVU Labex (ANR-11-LABX-0060), the  Spin(e) project (ANR-13-BS05-0005, \url{http://cosmicorigin.org}) and the A*MIDEX project (ANR-11-IDEX-0001-02) funded by the "Investissements d'Avenir" French government program managed by the ANR.
This publication makes use of Sloan Digital Sky Survey (SDSS) data. Funding for SDSS-III has been provided by the Alfred P. Sloan Foundation, the Participating Institutions, the National Science Foundation, and the U.S. Department of Energy Office of Science. The SDSS-III web site is \url{http://www.sdss3.org/}.
SDSS-III is managed by the Astrophysical Research Consortium for the Participating Institutions of the SDSS-III Collaboration including the University of Arizona, the Brazilian Participation Group, Brookhaven National Laboratory, Carnegie Mellon University, University of Florida, the French Participation Group, the German Participation Group, Harvard University, the Instituto de Astrofisica de Canarias, the Michigan State/Notre Dame/JINA Participation Group, Johns Hopkins University, Lawrence Berkeley National Laboratory, Max Planck Institute for Astrophysics, Max Planck Institute for Extraterrestrial Physics, New Mexico State University, New York University, Ohio State University, Pennsylvania State University, University of Portsmouth, Princeton University, the Spanish Participation Group, University of Tokyo, University of Utah, Vanderbilt University, University of Virginia, University of Washington, and Yale University.


\appendix

\section{SQL query}
\label{query}
We selected galaxies with spectroscopic redshifts from the Main Galaxy Sample of the DR12 by running the following SQL query on the CasJob website: \\[0.5cm]
{\bf SELECT} \\
   za.specObjID,za.bestObjID,za.class,za.subClass,za.z,za.zErr,\\
   po.objID,po.type,po.flags, po.ra,po.dec, \\
   ... \\
  (po.petroMag\_r  - po.extinction\_r) as dered\_petro\_r,\\
  ...\\
   zp.z as zphot, zp.zErr as dzphot,\\
zi.e\_bv\_sfd,zi.primtarget,zi.sectarget,zi.targettype,\\ zi.spectrotype, zi.subclass,  \\
... \\
{\bf INTO}    mydb.SDSS\_DR12 \\
{\bf FROM}  SpecObjAll za\\
{\bf JOIN}   PhotoObjAll  po  ON  (po.objID         = za.bestObjID)  \\
{\bf JOIN}   Photoz            zp   ON  (zp.objID          = za.bestObjID)  \\
{\bf JOIN}  galSpecInfo   zi    ON  (zi.SpecObjID = za.specObjID)\\
{\bf WHERE}\\
za.z $>$ 0 and za.zWarning=0  \\
and za.targetType ='SCIENCE' and za.survey='sdss' \\
and za.class='GALAXY' and zi.primtarget$\ge$64 \\
and po.clean=1 and po.insideMask=0 \\
and dered\_petro\_r$\le$17.8 \\

This results in a final sample of 516,546 galaxies. \\

\section{Detailed CNN architecture}
In this section we give details on our architecture (see Table \ref{table_reseau}) and on the computational time needed for the training steps, with one GTX Titan X card.
The time needed to pass forward and backward a batch composed of 128 galaxies through the network takes 0.21 second. So an epoch, which is the time needed to pass the entire dataset, takes 
about 14 minutes. We let the network converge in 120,000 iterations at most, which corresponds to 30 epochs. Therefore the training phase takes approximatively 7 hours. 

\label{archi}
\begin{table*}[t]
\setcounter{table}{0}
\renewcommand{\thetable}{A\arabic{table}}
\centering
\begin{tabular}{|c|c|c|c|c|}
\hline 
{\small{}Layer} & {\small{}Inputs} & {\small{}Kernel size} & {\small{}$h\times w$} & {\small{}\#feature maps}\tabularnewline
\hline 
{\small{}C1} & {\small{}input image} & {\small{}$5\times5$} & {\small{}$64\times64$} & {\small{}$64$}\tabularnewline
\hline 
{\small{}P1} & {\small{}C1} & {\small{}$2\times2$ (stride 2 pix)} & {\small{}$32\times32$} & {\small{}$64$}\tabularnewline
\hline 
\hline 
{\small{}C2, C3, C4} & {\small{}P1} & {\small{}$1\times1$, $1\times1$, $1\times1$ } & {\small{}$32\times32$, $32\times32$, $32\times32$} & {\small{}$48$, $48$, $48$}\tabularnewline
\hline 
{\small{}C5} & {\small{}P1} & {\small{}$1\times1$} & {\small{}$32\times32$} & {\small{}$64$}\tabularnewline
\hline 
{\small{}C6} & {\small{}C2} & {\small{}$3\times3$} & {\small{}$32\times32$} & {\small{}$64$}\tabularnewline
\hline 
{\small{}C7} & {\small{}C3} & {\small{}$5\times5$} & {\small{}$32\times32$} & {\small{}$64$}\tabularnewline
\hline 
{\small{}P2} & {\small{}C4} & {\small{}$2\times2$ (stride 1 pix)} & {\small{}$32\times32$} & {\small{}$64$}\tabularnewline
\hline 
{\small{}Co1} & {\small{}-} & {\small{}-} & {\small{}$32\times32$} & {\small{}$256$}\tabularnewline
\hline 
\hline 
{\small{}C8, C9, C10} & {\small{}Co1} & {\small{}$1\times1$, $1\times1$, $1\times1$ } & {\small{}$32\times32$, $32\times32$, $32\times32$} & {\small{}$64$, $64$, $64$}\tabularnewline
\hline 
{\small{}C11} & {\small{}Co1} & {\small{}$1\times1$} & {\small{}$32\times32$} & {\small{}$92$}\tabularnewline
\hline 
{\small{}C12} & {\small{}C8} & {\small{}$3\times3$} & {\small{}$32\times32$} & {\small{}$92$}\tabularnewline
\hline 
{\small{}C13} & {\small{}C9} & {\small{}$5\times5$} & {\small{}$32\times32$} & {\small{}$92$}\tabularnewline
\hline 
{\small{}P3} & {\small{}C10} & {\small{}$2\times2$ (stride 1 pix)} & {\small{}$32\times32$} & {\small{}$64$}\tabularnewline
\hline 
{\small{}Co2} & {\small{}-} & {\small{}-} & {\small{}$32\times32$} & {\small{}$340$}\tabularnewline
\hline 
\hline 
{\small{}P4} & {\small{}Co2} & {\small{}$2\times2$ (stride 2 pix)} & {\small{}$16\times16$} & {\small{}$340$}\tabularnewline
\hline 
{\small{}C14, C15, C16} & {\small{}P4} & {\small{}$1\times1$, $1\times1$, $1\times1$ } & {\small{}$16\times16$, $16\times16$, $16\times16$} & {\small{}$92$, $92$, $92$}\tabularnewline
\hline 
{\small{}C17} & {\small{}P4} & {\small{}$1\times1$} & {\small{}$16\times16$} & {\small{}$128$}\tabularnewline
\hline 
{\small{}C18} & {\small{}C14} & {\small{}$3\times3$} & {\small{}$16\times16$} & {\small{}$128$}\tabularnewline
\hline 
{\small{}C19} & {\small{}C15} & {\small{}$5\times5$} & {\small{}$16\times16$} & {\small{}$128$}\tabularnewline
\hline 
{\small{}P5} & {\small{}C16} & {\small{}$2\times2$ (stride 1 pix)} & {\small{}$16\times16$} & {\small{}$92$}\tabularnewline
\hline 
{\small{}Co3} & {\small{}-} & {\small{}-} & {\small{}$16\times16$} & {\small{}$476$}\tabularnewline
\hline 
\hline 
{\small{}C20, C21, C22} & {\small{}Co3} & {\small{}$1\times1$, $1\times1$, $1\times1$ } & {\small{}$16\times16$, $16\times16$, $16\times16$} & {\small{}$92$, $92$, $92$}\tabularnewline
\hline 
{\small{}C23} & {\small{}Co3} & {\small{}$1\times1$} & {\small{}$16\times16$} & {\small{}$128$}\tabularnewline
\hline 
{\small{}C24} & {\small{}C20} & {\small{}$3\times3$} & {\small{}$16\times16$} & {\small{}$128$}\tabularnewline
\hline 
{\small{}C25} & {\small{}C21} & {\small{}$5\times5$} & {\small{}$16\times16$} & {\small{}$128$}\tabularnewline
\hline 
{\small{}P6} & {\small{}C22} & {\small{}$2\times2$ (stride 1 pix)} & {\small{}$16\times16$} & {\small{}$92$}\tabularnewline
\hline 
{\small{}Co4} & {\small{}-} & {\small{}-} & {\small{}$16\times16$} & {\small{}$476$}\tabularnewline
\hline 
\hline 
{\small{}P7} & {\small{}Co4} & {\small{}$2\times2$ (stride 2 pix)} & {\small{}$8\times8$} & {\small{}$476$}\tabularnewline
\hline 
{\small{}C26, C27} & {\small{}P7} & {\small{}$1\times1$, $1\times1$} & {\small{}$8\times8$, $8\times8$} & {\small{}$92,$$92$}\tabularnewline
\hline 
{\small{}C28} & {\small{}P7} & {\small{}$1\times1$} & {\small{}$8\times8$} & {\small{}$128$}\tabularnewline
\hline 
{\small{}C29} & {\small{}C26} & {\small{}$3\times3$} & {\small{}$8\times8$} & {\small{}$128$}\tabularnewline
\hline 
{\small{}P8} & {\small{}C27} & {\small{}$2\times2$ (stride 1 pix)} & {\small{}$8\times8$} & {\small{}$92$}\tabularnewline
\hline 
{\small{}Co5} & {\small{}-} & {\small{}-} & {\small{}$8\times8$} & {\small{}$348$}\tabularnewline
\hline 
\hline 
{\small{}FC1, FC2} & {\small{}Co5, FC1} & {\small{}-} & {\small{}-} & {\small{}$1024$, $1024$}\tabularnewline
\hline 
\end{tabular}
\caption{Characteristics of each layer of the CNN architecture: name of the layer, input layer, size of the convolution kernel (in pixels), size (height$\times$width in pixels) and number of the resulting feature maps. }
\label{table_reseau}
\end{table*}

\section{Impact of image quality on the CNN performance}
\label{performance}
\begin{figure*}
\includegraphics[width=6.5cm]{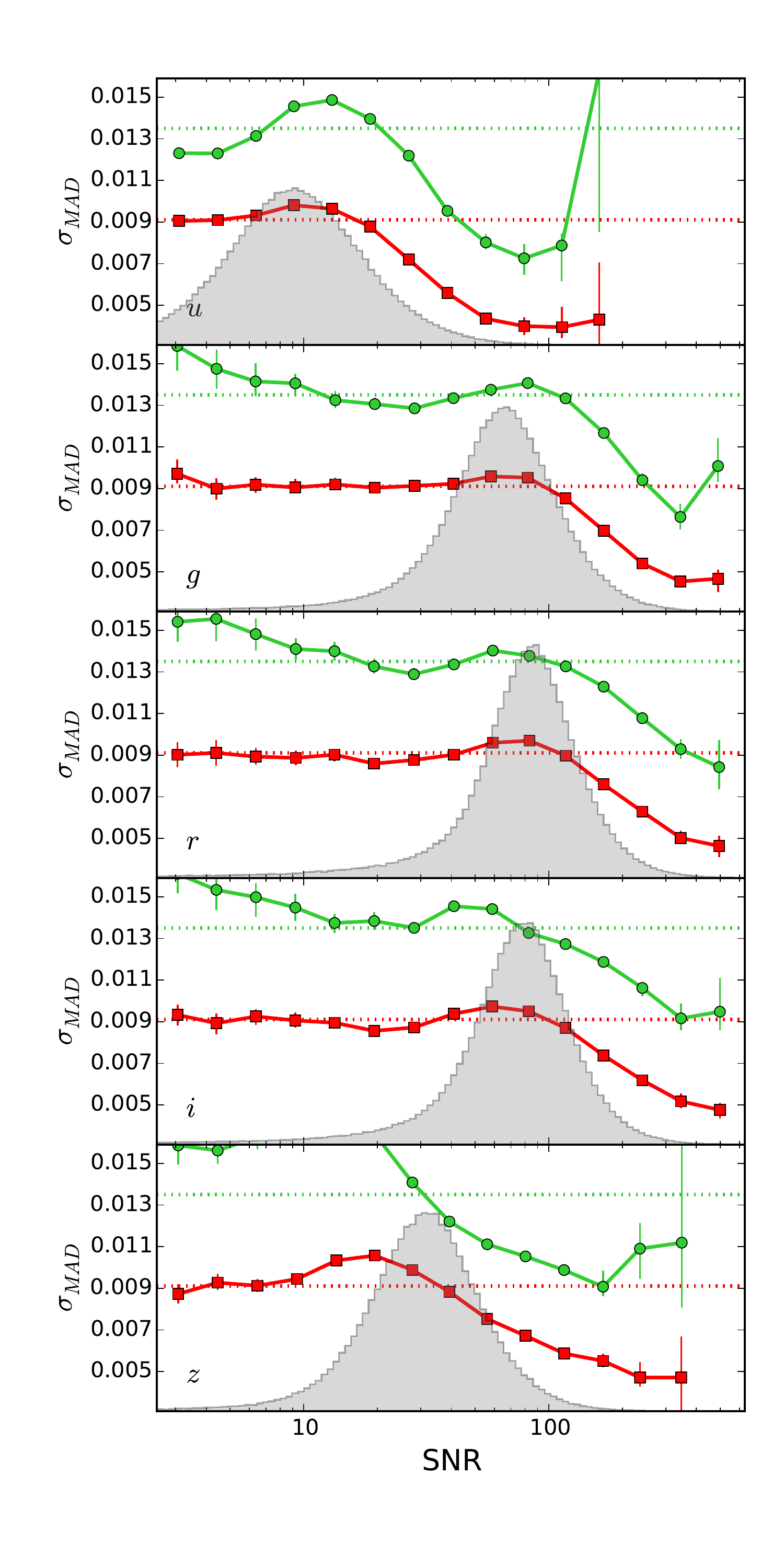}
\hspace{-0.4cm}
\includegraphics[width=6.5cm]{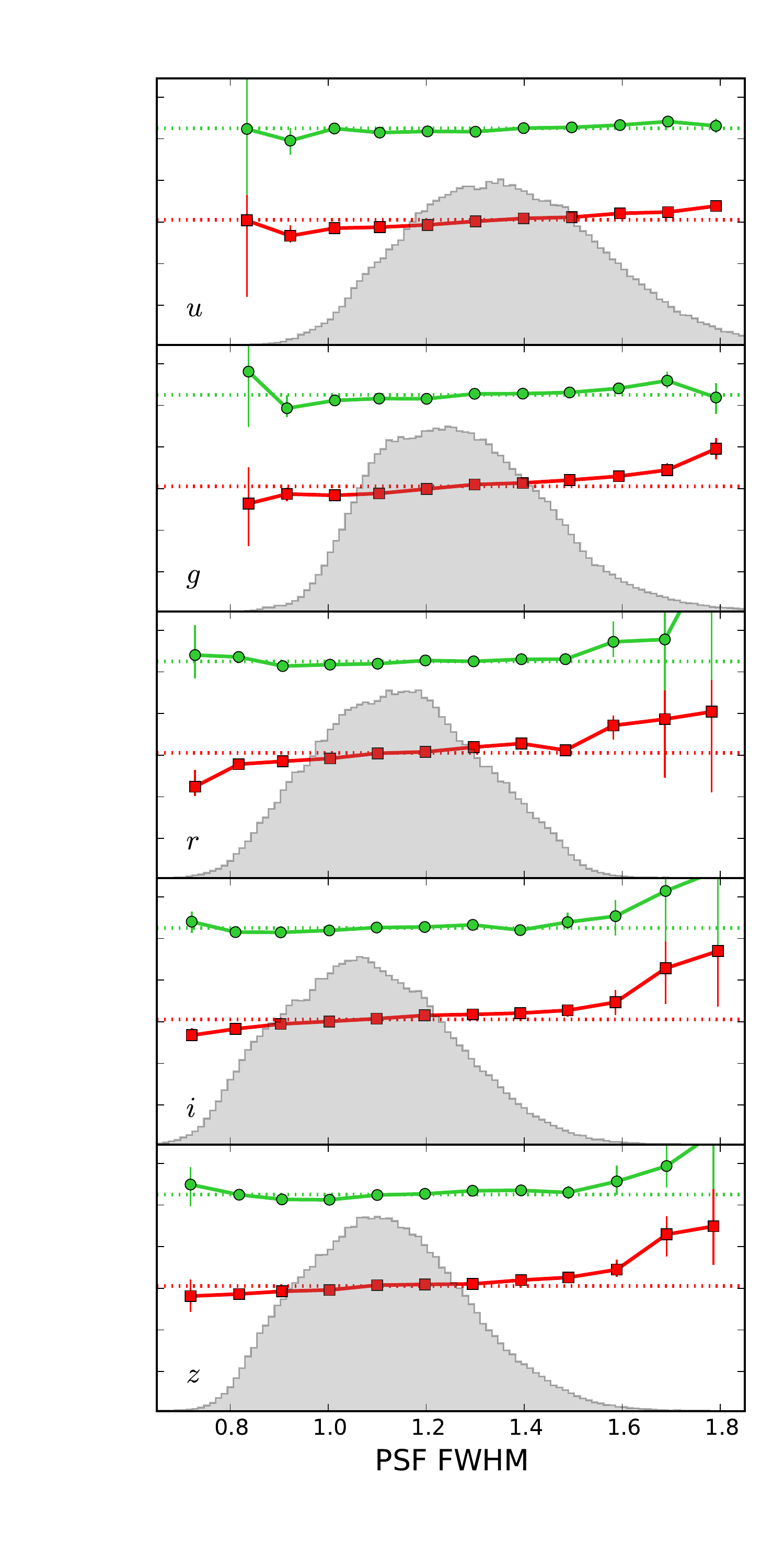}
\hspace{-0.4cm}
\includegraphics[width=6.5cm]{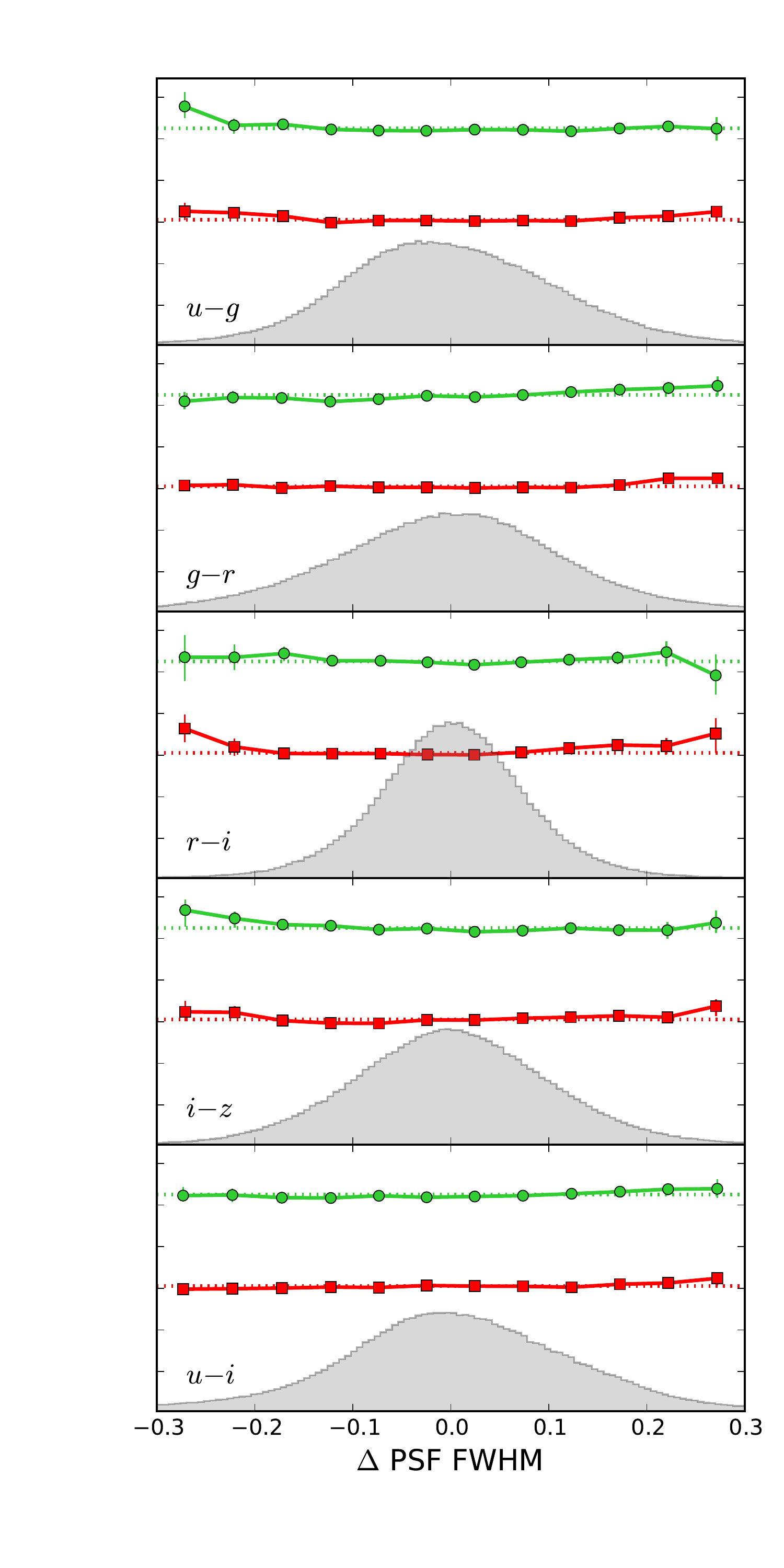}

\includegraphics[width=6.5cm]{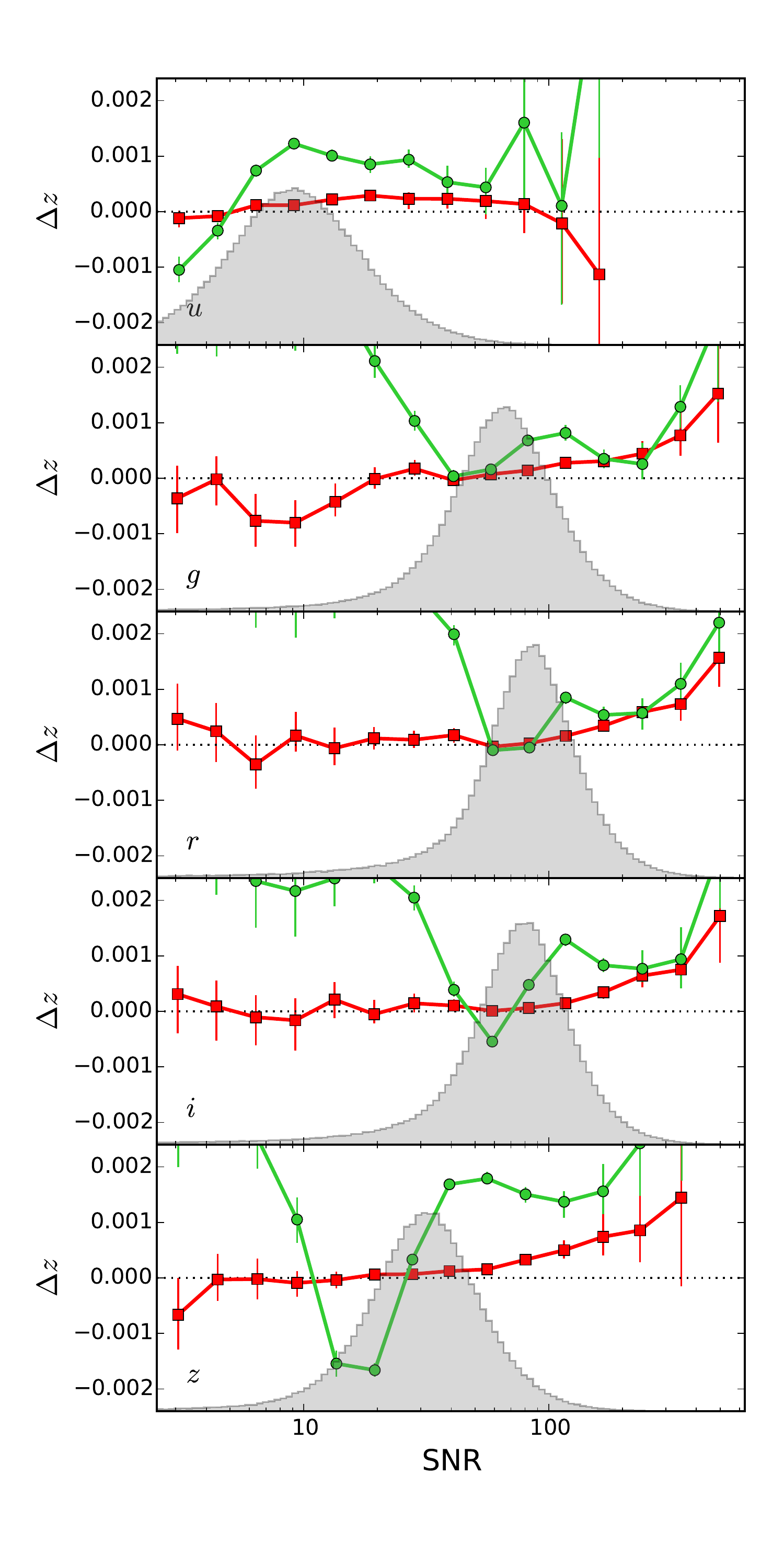}
\hspace{-0.4cm}
\includegraphics[width=6.5cm]{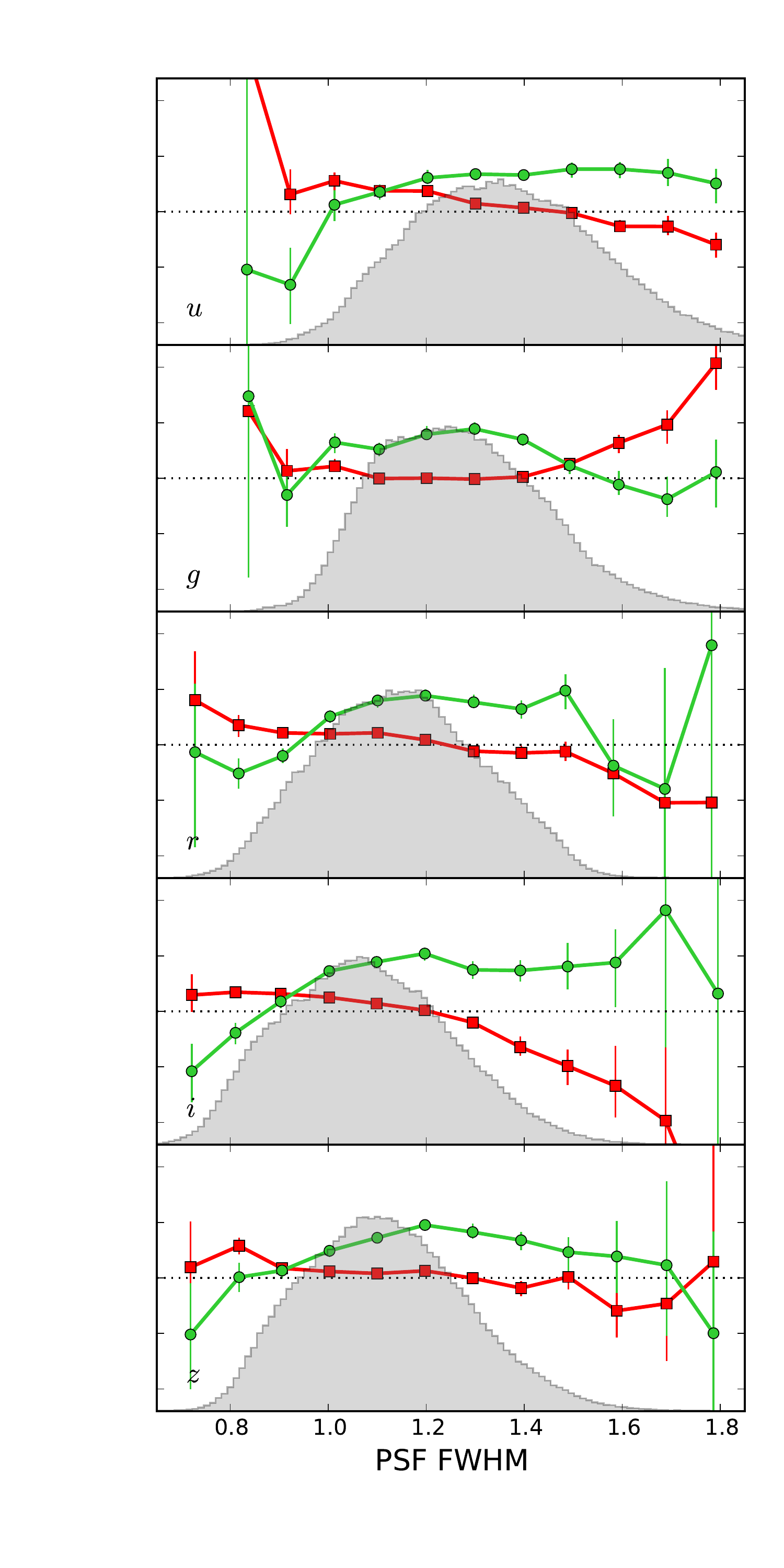}
\hspace{-0.4cm}
\includegraphics[width=6.5cm]{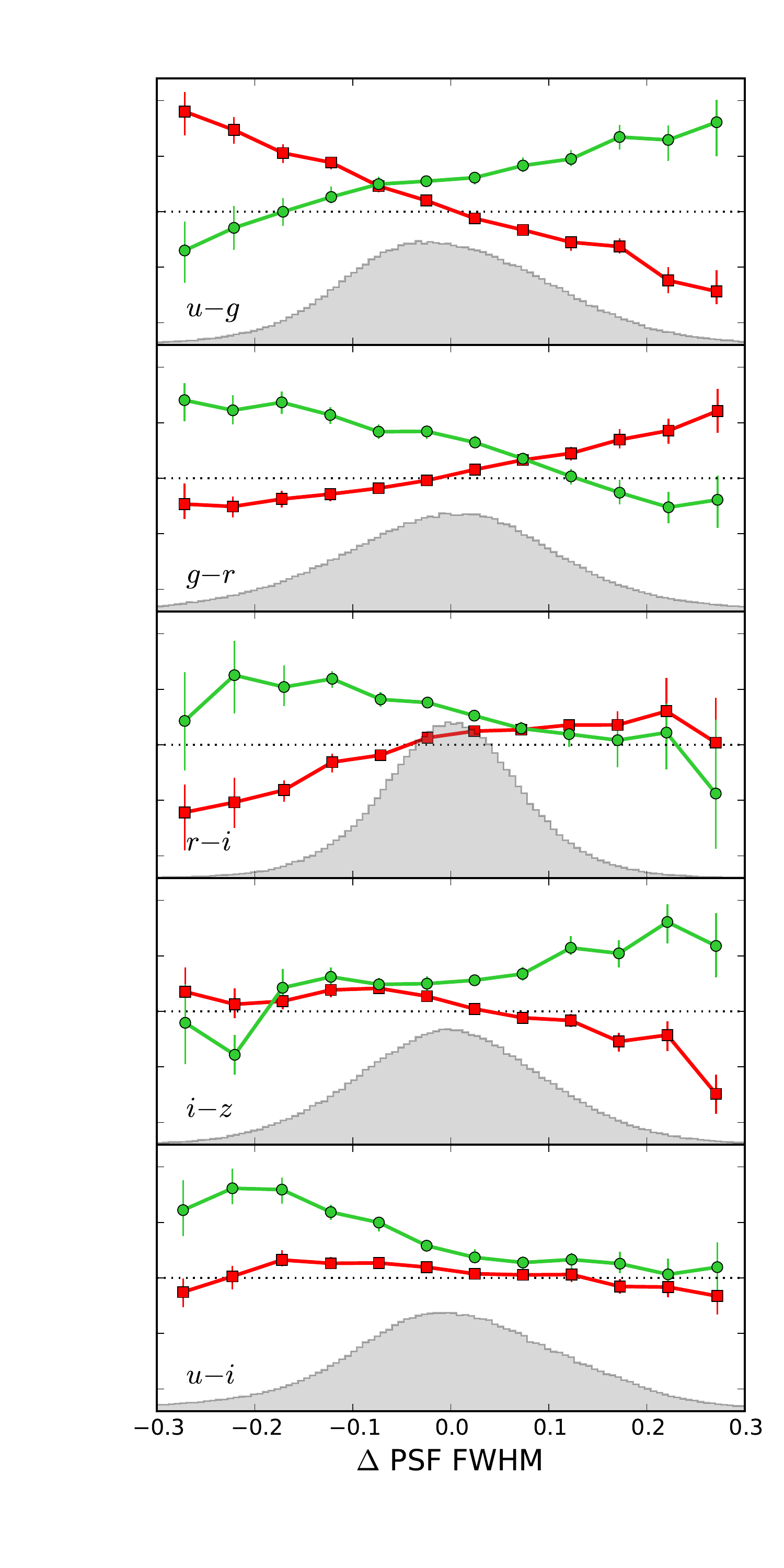}
\caption{$\sigma_{\rm MAD}$ and bias as a function of signal-to-noise ratio (SNR=1.086/petroMagErr) and PSF FWHM in the 5 bands, and of PSF FWHM offset between 2 bands (recentered at the mean value). The CNN results are shown in red, the B16 results in green.
\label{performances appendix}}
\end{figure*}

\bibliographystyle{aa}
\bibliography{aa.bib} 
\end{document}